\documentclass[aps,prd,preprint,onecolumn,nofootinbib,a4paper]{revtex4-1}

\usepackage{amsmath,amssymb,amsfonts}
\usepackage[colorlinks=true,urlcolor=blue,linkcolor=blue,citecolor=red]{hyperref}
\usepackage{bm,soul}
\usepackage{xcolor}
\usepackage{longtable}
\usepackage{amsmath}
\usepackage{mathtools}
\usepackage[utf8]{inputenc}

\def\be{\begin{equation}}
\def\ee{\end{equation}}
\def\ba{\begin{eqnarray}}
\def\ea{\end{eqnarray}}
\newcommand{\D}[0]{\mathrm{D}}
\def\onehalf{{\textstyle{\frac{1}{2}}}}
\def\onefourth{{\textstyle{\frac{1}{4}}}}

\def\Gammaw{{\stackrel{\mbox{\tiny$\bullet$}}{\Gamma}}{}}
\def\gammaw{{\stackrel{\mbox{\tiny$\bullet$}}{\gamma}}{}}
\def\omegaw{{\stackrel{\mbox{\tiny$\bullet$}}{\omega}}{}}

\def\Rw{{\stackrel{\mbox{\tiny$~\bullet$}}{R}}{}}

\def\jw{{\stackrel{\mbox{\tiny$~\bullet$}}{\jmath}}{}}

\def\tw{{\stackrel{\mbox{\tiny$~\bullet$}}{t}}{}}
\def\L{{\mathcal L}{}}
\def\Lw{{\stackrel{\mbox{\tiny$~\bullet$}}{\L}}{}}
\def\Tw{{\stackrel{\mbox{\tiny$\bullet$}}{T}}{}}
\def\Bw{{\stackrel{\mbox{\tiny$\bullet$}}{B}}{}}
\def\Sigmaw{{\stackrel{\mbox{\tiny$\bullet$}}{\Sigma}}{}}
\def\Upsilonw{{\stackrel{\mbox{\tiny$\bullet$}}{\Upsilon}}{}}
\def\Piw{{\stackrel{\mbox{\tiny$\bullet$}}{\Pi}}{}}
\def\Hw{{\stackrel{\mbox{\tiny$\bullet$}}{H}}{}}

\def\Kw{{\stackrel{\mbox{\tiny$\;\bullet$}}{K}}{}}
\def\Aw{{\stackrel{\mbox{\tiny$\bullet$}}{\omega}}{}}
\def\nablaw{{\stackrel{\mbox{\tiny$\bullet$}}{\nabla}}{}}
\def\Dw{{\stackrel{\mbox{\tiny$~\bullet$}}{\mathcal D}}{}}

\def\Sw{{\stackrel{\mbox{\tiny$~\bullet$}}{\mathcal S}}{}}
\def\Sigmaw{{\stackrel{\mbox{\tiny$\bullet$}}{\Sigma}}{}}

\def\sw{{\stackrel{\mbox{\tiny$~\bullet$}}{S}}{}}

\def\omw{{\stackrel{\mbox{\tiny$~\bullet$}}{\omega}}{}}
\def\actionw{{\stackrel{\mbox{\tiny~$\bullet$}}{\mathcal S}}{}}

\def\Gammabol{{\stackrel{\mbox{\tiny$\circ$}}{\Gamma}}{}}{}

\def\Abol{{\stackrel{\mbox{\tiny$\circ$}}{\omega}}{}}
\def\Gbol{{\stackrel{\circ}{G}}{}}

\def\Rbol{{\stackrel{\mbox{\tiny$~\circ$}}{R}}{}}
\def\Lbol{{\stackrel{~\circ}{\mathcal L}}{}}

\def\Dbol{{\stackrel{\mbox{\tiny$\circ$}}{\mathcal D}}{}}

\def\omegabol{{\stackrel{\mbox{\tiny$\circ$}}{\omega}}{}}

\def\tref{h_{(\rm r)}}

\begin{document}


\title{Teleparallel Theories of Gravity: Illuminating a Fully Invariant Approach}

\author{M. \surname{Kr\v{s}\v{s}\'{a}k}}
\email{martin.krssak@ut.ee}
\affiliation{Center for Gravitation and Cosmology, College of Physical Sciences
	and Technology, Yangzhou University, Yangzhou 225009, China\\}
\affiliation{Laboratory of Theoretical Physics, Institute of Physics, University of Tartu, W. Ostwaldi 1, Tartu 50411, Estonia }

\author{R. J. \surname{van den Hoogen}}
\email{rvandenh@stfx.ca}
\affiliation{Department of Mathematics and Statistics, St. Francis Xavier University, P.O. Box 5000, Antigonish, N.S., B2G 2W5, Canada}

\author{J. G. \surname{Pereira}}
\email{jg.pereira@unesp.br}
\affiliation{Instituto de F\'{i}sica Te\'{o}rica, Universidade Estadual Paulista, R. Dr. Bento Teobaldo Ferraz 271, 01140-070, S\~{a}o Paulo, Brazil}

\author{C. G. \surname{B\"{o}hmer}}
\email{c.boehmer@ucl.ac.uk}
\affiliation{Department of Mathematics, University College London, Gower Street, London WC1E 6BT, United Kingdom}

\author{A. A. \surname{Coley}}
\email{aac@mathstat.dal.ca}
\affiliation{Department of Mathematics and Statistics, Dalhousie University, 6316 Coburg Road, P.O. BOX 15000, Halifax, N.S., B3H 4R2, Canada}

\date{\today}


\newpage
\begin{abstract}

Teleparallel gravity and its popular generalization $f(T)$ gravity can be formulated as fully invariant (under both coordinate transformations and local Lorentz transformations) theories of gravity.  Several misconceptions about teleparallel gravity and its generalizations can be found in the literature,
especially regarding their local Lorentz invariance. We describe how these misunderstandings may have arisen and attempt to clarify the situation.  In particular, the central point of confusion in the literature appears to be related to the inertial spin connection in teleparallel gravity models. While inertial spin connections are commonplace in special relativity, and not something inherent to teleparallel gravity, the role of the inertial spin connection in removing the spurious inertial effects within a given frame of reference is emphasized here. The careful consideration of the inertial spin connection leads to the construction of a fully invariant theory of teleparallel gravity and its generalizations.  Indeed, it is the nature of the spin connection that differentiates the relationship between what have been called \emph{good tetrads} and \emph{bad tetrads} and clearly shows that, in principle, any tetrad can be utilized. The field equations for the fully invariant formulation of teleparallel gravity and its generalizations are presented and a number of examples using different assumptions on the frame and spin connection are displayed to illustrate the covariant procedure. Various modified teleparallel gravity models are also briefly reviewed.
\end{abstract}

\maketitle

\tableofcontents
\thispagestyle{empty}
\newpage

\section{Introduction}\label{sec:Intro}

Although Einstein's General Theory of Relativity (GR) is well studied and tested in many settings~\cite{Will:2014kxa}, alternative theories of gravity continue to be of considerable interest~\cite{Nojiri:2010wj,Clifton:2011jh,Capozziello:2011et,Nojiri:2017ncd,Heisenberg:2018vsk}.  Potential explanations for dark energy and dark matter within the current cosmological paradigm based on general relativity may be investigated using an alternative theory of gravity, rather than changing the matter content of the theory. Furthermore, it is possible that the problems of finding a quantum theory of gravity may be resolved within a theory of gravity that is not general relativity. Therefore, it becomes necessary to challenge our assumptions and assess whether an alternative theory of gravity will lead to different results.

One class of alternative theories of gravity assumes that the  motion in  the gravitational field is no longer geometrized, as in general relativity, but is encoded in a dynamic gravitational force, as in teleparallel gravity. More specifically,  in general relativity the gravitational interaction is realized via the curvature of a zero torsion Lorentz connection, which is used to geometrize the interaction; this means that the motion of  a free-falling particle in the gravitational field can be viewed as an inertial motion in the curved spacetime and hence  gravity can be viewed as a purely geometric effect.  On the other hand, in teleparallel gravity the gravitational interaction is an effect of the torsion of a zero curvature Lorentz connection. Torsion in this case acts as a force, which similarly to the Lorentz force equation of electromagnetism, appears as an effective force term on the right-hand side of the equation of motion of a free-falling particle. We see in this way that, even though torsion has a well defined geometrical meaning, this geometrical meaning is not relevant for the teleparallel description of the gravitational interaction. Interestingly, teleparallel gravity and general relativity   are found to be completely equivalent theories. For this reason one generally refers to it as the \emph{Teleparallel Equivalent of General Relativity (TEGR)}. Although equivalent, however, they are conceptually quite different. For example, in contrast to general relativity, teleparallel gravity is nicely motivated within a gauge theory context and can be beautifully framed as the gauge theory for the translation group~\cite{Aldrovandi_Pereira2013}. In fact, like all other gauge theories, its Lagrangian density is quadratic in the torsion tensor, the field strength of the theory. The notions of frame and inertial spin connection are presented in Section~\ref{sec:frames}. The fundamentals of teleparallel gravity are described in Sections~\ref{secBasics} and~\ref{sec:LagrangianTeleGrav}.

The geometrical setting of any gravitational theory is the tangent bundle, in which spacetime is the base space and the tangent space at each point of the base space (also known as internal space) is the fiber of the bundle. Spacetime is assumed to be a metric spacetime with a general metric $g_{\mu \nu}$. The tangent space, on the other hand, is by definition a Minkowski spacetime with tangent space metric $\eta_{ab}$. Since spacetime and the fibers are both four-dimensional spacetimes, the bundle is said to be soldered. This means that the metrics $g_{\mu \nu}$ and $\eta_{ab}$ are related by
\[
g_{\mu \nu} = \eta_{ab} \, h^a{}_\mu h^b{}_\nu \, ,
\]
with $h^a{}_\mu$ being the tetrad field, the components of the solder 1-form. It should be noted that this geometrical structure is always present, independent of any prior assumptions.

Teleparallel gravity, a gauge theory for the translation group, is built on this geometrical structure. Gauge transformations are defined as local translations in the tangent Minkowski spacetime, the fiber of the bundle. Of course, like any other relativistic theory, it must also be invariant under both general coordinate transformations and local Lorentz transformations. Whereas the former is performed in spacetime, the latter is performed in the tangent space.

\emph{Local} Lorentz transformations define different classes of frames, each one characterized by different inertial effects represented by a purely inertial connection (which we define later). Within each class, the infinitely many equivalent frames are related by a \emph{global} Lorentz transformation. In the class of frames in which no inertial effects are present, the inertial Lorentz connection is naturally zero.\footnote{In the presence of gravitation, these frames are called ``proper frames''. In the absence of gravitation they reduce to the class of inertial frames of special relativity.} In all other classes of frames, however, their inertial spin connection will be non-vanishing.

Although they produce physical effects and have energy and momentum, inertial effects cannot be interpreted as a field in the usual sense of classical field theory. For example, in TEGR there are no field equations whose solutions could yield the inertial Lorentz connection.\footnote{The situation is more subtle in modified theories such as $f(T)$, and we will discuss this later (e.g., see the comments in the Final Remarks).}
(Indeed, neither the field equations of teleparallel gravity nor the field equations of general relativity are able to determine this.) Since the use of the correct inertial Lorentz connection is crucial for the Lorentz symmetry of any relativistic symmetry, it is then necessary to resort to a different method for retrieving the inertial Lorentz connection associated to a general frame. Such a method is presented in detail in Section~\ref{seccon}, and some concrete examples are discussed in Section~\ref{secExample}.

It is important to remark that in the usual metric formulation of general relativity  no frame needs to be specified. In the tetrad formulation of  general relativity, we do not face the problem of specifying the Lorentz connection because  the Levi-Civita spin connection of general relativity can be fully expressed in terms of the dynamical tetrad and hence can be  eliminated from the theory. Furthermore, the Levi-Civita connection includes both gravitational and inertial effects, unlike teleparallel gravity where gravitation is represented by a translational gauge potential and inertial effects are represented by an inertial spin connection. In contrast to general relativity, therefore, the question of specifying the inertial spin connection is part of the process of finding solutions for the teleparallel field equations.

Influenced perhaps by general relativity, in which one does not need to carefully consider the inertial spin connection, many authors have never scrutinized it when working in the context of teleparallel gravity. As a consequence much of the work on the teleparallel gravity was done using the assumption that the spin connection can be always chosen to be zero, and the only variable being the frame field, or tetrad.  Of course, owing to the disregard of the inertial spin connection, the resulting theory is not invariant under local Lorentz transformations. This non-covariant version of the theory was then later nick-named ``pure tetrad teleparallel gravity" due to the fact that the tetrad appears as the only field variable~\cite{Krssak_Pereira2015}.

In spite of this lack of Lorentz invariance, since the solutions of the field equations of teleparallel gravity are independent of the inertial spin connection, the solutions provided by the ``pure tetrad teleparallel gravity'' coincide with the solutions provided by the locally Lorentz invariant formulation of teleparallel gravity. Nevertheless, apart from this property, some other conclusions obtained from this theory, including the Lorentz non-invariance, are different from those obtained from the locally Lorentz  invariant teleparallel gravity theory. A discussion on ``pure tetrad teleparallel gravity'', as well as on its differences in relation to teleparallel gravity itself, is presented in Section~\ref{secPT}.

Recently there have been many proposals to generalize teleparallel gravity. This was motivated by the example of $f(\Rbol)$ gravity, where the Lagrangian density is generalized from $\Rbol\to f(\Rbol)$, where $\Rbol$ is the Ricci scalar of the Levi-Civita connection (see~\cite{Sotiriou:2008rp} and references within for an overview of $f(R)$ gravity\footnote{The notation used in this paper is summarized in Table I below.}).
Similarly, it was suggested  to generalize the Lagrangian density of teleparallel gravity from $\Tw\to f(\Tw)$, where $\Tw$ is the so-called torsion scalar with respect to the teleparallel connection that we will define later (see~\cite{Cai_2015} and references within for an overview of $f(T)$ teleparallel gravity). Unlike the generalization of general relativity, in which the resulting field equations are no longer second order in derivatives, in $f(T)$ teleparallel gravity the field equations continue to be second order. Unfortunately, there appears to be some confusion in the literature with regards to the viability of $f(T)$ teleparallel gravity and, in particular, to its invariance under local Lorentz transformations~\cite{Li_Sotiriou_Barrow2010,Sotiriou_Li_Barrow2011}. In the generalization of the type  $\Rbol \to f(\Rbol)$, since the Ricci scalar is built from the metric, which is Lorentz invariant by definition, the Ricci scalar $\Rbol$ is naturally Lorentz invariant too, and therefore $f(R)$ gravity is also Lorentz invariant. In generalizations of the type $\Tw\to f(\Tw)$, since $\Tw$ is a combination of scalar invariants of the torsion, and since the torsion tensor is a Lorentz covariant object, the scalar $\Tw$ is Lorentz invariant, and consequently so is $f(T)$. It appears that the source of confusion is in the use of the ``pure tetrad teleparallel gravity", which is not invariant under local Lorentz transformations. In this case, the non-invariance of the ``pure tetrad teleparallel gravity" will of course propagate to the modified $f(T)$ models.

Fortunately, provided the inertial spin connection is appropriately taken into account, teleparallel gravity can be seen to be fully invariant under local Lorentz transformations. In this case, provided the same care is used, a fully covariant $f(T)$ theory can be obtained~\cite{Krssak_Saridakis2015}. Details of this construction are presented in Sec.~\ref{secfT}. Using an analogous procedure, it is possible to extend teleparallel gravity to other modified gravity models. In Sec.~\ref{sec:Other-Modified} we illustrate this possibility with a number of modified teleparallel theories of gravity, including new general relativity~\cite{Hayashi_Shirafuji1979}, conformal teleparallel gravity~\cite{Maluf:2011kf} and $f(T,B)$ gravity~\cite{Bahamonde:2015zma}.

\subsection{Notation}

The notation used when formulating teleparallel theories of gravity resembles that used in general relativity. However, it is necessary to introduce some additional symbols for the various quantities which naturally arise in this framework. Sometimes this notation is identical to the notation used in general relativity but actually denotes a slightly different object. In order to help the reader navigate this notational quagmire, we present a list of symbols used throughout this work.

\setlength{\tabcolsep}{1ex}
\begin{longtable}[c]{| c | @{\qquad\qquad}l |}
  \hline
  \multicolumn{2}{|c|}{Table~\ref{long}}\\
  \hline
  Symbol & Description \\
  \hline
  \endfirsthead

  \hline
  \multicolumn{2}{|c|}{Continuation of Table~\ref{long}}\\
  \hline
  Symbol & Description \\
  \hline
  \endhead

  \hline
  \endfoot
  \endlastfoot
  $\mu,\nu,\dots$ & coordinate indices \\
  $a,b,\dots$ & tangent space indices \\
  $x^\mu$ & space-time coordinates \\
  $e_a$ & trivial frame fields \\
  $e^a$ & trivial coframe one-forms \\
  $e_a{}^\mu$ & trivial frame field components \\
  $e^a{}_\mu$ & trivial coframe one-form components \\
  $f^c{}_{ab}$ & coefficients of anholonomy \\
  $\eta_{\mu\nu}$ & Minkowski spacetime metric \\
  $g_{\mu\nu}$ & arbitrary spacetime metric \\
  $\eta_{ab}$ & Minkowski tangent space metric \\
  $\Lambda^a{}_b(x)$ & local Lorentz transformation \\
  $\epsilon_{ab}$ & infinitesimal Lorentz transformation \\
  $\Aw^a{}_{b\mu}$ & teleparallel spin connection\\
  $\Dw_{\mu}$ & covariant derivative associated with $\Aw^a{}_{b\mu}$ \\
  $\omega^a{}_{b\mu}$ & general spin connection \\
  $\Rw^a{}_{b\mu\nu}$ & Riemann curvature tensor of $\Aw^a{}_{b\mu}$ \\
  $\Tw^a{}_{\mu\nu}$ & torsion tensor of $\Aw^a{}_{b\mu}$ \\
  $\Tw^\mu$ & torsion vector defined by $\Tw^{\nu\mu}{}_{\nu}$ \\
  $u^a$ & anholonomic 4-velocity \\
  $u^\mu$ & holonomic 4-velocity \\
  $d\sigma$ & Minkowski interval \\
  $ds$ & arbitrary interval \\
  $\gammaw^{\rho}{}_{\mu\nu}$ & holonomic connection associated with $\Aw^a{}_{b\mu}$ \\
  $\nablaw_{\mu}$ & covariant derivative associated with $\gammaw^{\rho}{}_{\mu\nu}$ \\
  $\varepsilon^a(x^{\mu})$ & local tangent space translation \\
  $\delta_\varepsilon$ & change of quantity under translation \\
  $B^a{}_{\mu}$ & gauge potential one-form components \\
  $h_{\mu}$ & gauge covariant derivative \\
  $h^a$ &   non-trivial frame field\\
  $h_a$ &   non-trivial coframe field \\
  $h^a{}_{\mu}$ & non-trivial frame field components \\
  $h_a{}^{\mu}$ & non-trivial coframe field components \\
  $h_{(\rm r)}^a{}_{\mu}$ & reference tetrad field\\
  $h$ & determinant of the tetrad \\
  $\Gammaw^{\rho}{}_{\mu\nu}$ & teleparallel linear ({Weitzenb\"ock}) connection \\
  $\Kw^c{}_{ba}$ & contortion tensor \\
  $\Tw$ & torsion scalar \\
  $\Bw$ & boundary term \\
  $\Lw$ & Lagrangian density of teleparallel gravity\\
  $\actionw$ & action of teleparallel gravity\\
  $\kappa = 8\pi G$ & gravitational coupling constant ($c=1$)\\
  $E_a{}^\rho$ & Euler-Lagrange expression \\
  $\sw_a{}^{\rho\sigma}$ & superpotential \\
  $\jw_a{}^{\rho}$ & gauge current or energy-momentum pseudo-current \\
  $\Theta_a{}^\rho$ & matter energy-momentum tensor \\
  $\Sigmaw_a{}^{\rho}$ & gravitational energy-momentum tensor \\
  $\tw_{\mu}{}^{\rho}$ & energy-momentum pseudo-tensor \\
  $\Aw^\mu$ & quantity defined by $\Aw^a{}_{b\nu} h_a{}^\nu h_s{}^{\mu} \eta^{bs}$ \\
  $\omegabol^c{}_{b\mu}$ & spin connection of general relativity \\
  $\Gammabol^\lambda_{\phantom{\mu}\mu\nu}$ & Christoffel symbol or general relativity connection \\
  $\Rbol^{\mu}{}_{\nu}$ & Ricci tensor of general relativity \\
  $\Rbol$ & Ricci scalar of general relativity \\
  $\Lbol$ & Lagrangian density of general relativity\\
  $\Lw_f$ & Lagrangian density of $f(T)$ gravity\\
  $v_{\mu}, a_{\mu}, t_{\lambda\mu\nu}$ & irreducible pieces of the torsion tensor \\
  $T_{\rm vec}, T_{\rm ax},T_{\rm ten}$ & squares of irreducible torsion pieces
  \\  \hline
  \caption{Notation employed in the description of teleparallel gravity, general relativity and other gravitational theories used in this paper.\label{long}}\\
\end{longtable}


\section{Spin Connection in Special Relativity}\label{sec:frames}

\subsection{Trivial frames and bundles}

Trivial frames, or tetrads, represent observers in special relativity and hence exist only in the absence of gravity. They  are denoted  here by $\{e_{a}\}$ and $\{e^{a}\}$, and are general linear bases on the Minkowski spacetime manifold, satisfying the relation
\begin{equation}
e^{a}(e_b) = \delta^a_b.
\label{OrtoLiFra}
\end{equation}
The whole set of such bases constitutes the \emph{bundle of linear frames}. A frame field provides, at each point $p$ of spacetime, a basis for the vectors on the tangent space. Of course, on the common domains they are defined, each member of a given basis can be written in terms of the members of any other. For example,
\begin{equation}
e_a = e_a{}^\mu \, \partial_\mu \quad \mbox{and} \quad e^{a} = e^{a}{}_\mu \, dx^\mu,
\end{equation}
and conversely,
\begin{equation}
\partial_\mu = e^a{}_\mu \, e_a \quad \mbox{and} \quad dx^\mu = e_{a}{}^\mu \, e^{a}.
\label{eq:partialmu}
\end{equation}
On account of the orthogonality condition \eqref{OrtoLiFra}, the frame components satisfy
\begin{equation}
e^{a}{}_{\mu} e_{a}{}^{\nu} = \delta_{\mu}^{\nu} \quad \mbox{and} \quad
e^{a}{}_{\mu} e_{b}{}^{\mu} = \delta^{a}_{b}.
\label{eq:frameprops1}
\end{equation}
Notice that these frames, and their bundles, are constitutive parts of spacetime: they are present as soon as spacetime is taken to be a differentiable manifold \cite{Aldrovandi_Pereira2016}.

A general linear basis $\{e_{a}\}$ satisfies the commutation relation
\begin{equation}
[e_{a}, e_{b}] = f^{c}{}_{a b} \; e_{c},
\label{eq:comtable0}
\end{equation}
with $f^{c}{}_{a b}$ the so--called coefficients of anholonomy, which are functions of the spacetime points.
The dual expression of the commutation relation above is the Cartan structure equation
\begin{equation}
d e^{c} = -\, \onehalf \, f^{c}{}_{a b}\, e^{a} \wedge e^{b} = \onehalf \,
(\partial_\mu e^c{}_\nu - \partial_\nu e^c{}_\mu)\, dx^\mu \wedge dx^\nu.
\label{eq:dualcomtable0}
\end{equation}
The coefficient of anholonomy represent the curls of the basis members:
\begin{equation}
f^c{}_{a b} = e_a{}^{\mu} e_b{}^{\nu} (\partial_\nu
e^c{}_{\mu} - \partial_\mu e^c{}_{\nu} ).
\label{fcab0}
\end{equation}
A special class of frames is that of inertial frames, denoted $e'_a$, for which
\begin{equation}
f'^{a}{}_{cd} = 0.
\label{fcabinertial}
\end{equation}
Notice that $f'^{c}{}_{a b}$ = $0$ means that $d e'^{a} = 0$ which, in turn, implies that $e'^{a}$ is a closed differential form and, consequently, locally exact: $e'^{a}$ = $dx'^a$ for some $x'^a$. The basis $\{e'^{a}\}$ is then said to be integrable, or \emph{holonomic}. Of course, all coordinate bases are holonomic. This is not a local property in the sense that it is valid everywhere for frames belonging to this inertial class.

Consider now the Minkowski spacetime metric written in a holonomic basis $\{d x^{\mu}\}$. When $\{x^\mu \}$ represents a set of Cartesian coordinates, it has the form
\begin{equation}
\eta_{\mu \nu} = \mathrm{diag}(+1,-1,-1,-1).
\label{eq:etaofMinkoST}
\end{equation}
In any other coordinates, $\eta_{\mu \nu}$ will be a function of the spacetime coordinates. The linear frame $e_{a} = e_{a}{}^{\mu} \, {\partial_{\mu}}$ provides a relation between the tangent--space metric
$\eta_{a b}$ and the spacetime metric $\eta_{\mu \nu}$, given by
\begin{equation}
\eta_{a b} = {\eta}_{\mu \nu} \, e_{a}{}^{\mu} e_{b}{}^{\nu}
\label{gtoeta}
\end{equation}
with the inverse given by
\begin{equation}
{\eta}_{\mu \nu} = \eta_{a b} \, e^{a}{}_{\mu} e^{b}{}_{\nu}.
\label{eq:tettomet0}
\end{equation}
Independent of whether $e_{a}$ is holonomic or not, or equivalently, whether they are inertial or not, they always relate the tangent Minkowski space to a Minkowski spacetime. These are the frames appearing in special relativity, which are usually called trivial frames, or trivial tetrads.

\subsection{Spin connections and inertial effects}\label{sec:InerEff}

In special relativity, Lorentz connections represent inertial effects present in a given frame. In the class of inertial frames, for example, where these effects are absent, the Lorentz connection vanishes identically. To see how an inertial Lorentz connection shows up, let us consider an inertial frame $e'^a{}_\mu$ written in a general coordinate system $\{x^\mu\}$, in which case it has the holonomic form
\begin{equation}
e'^a{}_\mu = \partial_\mu x'^a
\label{frame0}
\end{equation}
with $x'^a$ a point--dependent Lorentz vector: $x'^a = x'^a(x^\mu)$. Under a local Lorentz transformation,
\begin{equation}
x^a = \Lambda^a{}_b(x) \, x'^b,
\end{equation}
the holonomic frame \eqref{frame0} transforms into the new frame
\begin{equation}
e^a{}_\mu = \Lambda^a{}_b(x) \, e'^b{}_\mu.
\label{LoreTrans-e}
\end{equation}
As a simple computation shows, it has the explicit form
\begin{equation}
e^a{}_\mu = \partial_\mu x^a + \Aw^a{}_{b \mu} \, x^b \equiv \Dw_\mu x^a
\label{InertiaTetrad}
\end{equation}
where
\begin{equation}
\Aw^a{}_{b \mu} = \Lambda^a{}_e(x) \, \partial_\mu \Lambda_b{}^e(x)
\label{InerConn}
\end{equation}
is a Lorentz connection that represents the inertial effects present in the Lorentz--rotated frame $e^a{}_\mu$, and $\Dw_\mu$ is the associated covariant derivative. Recalling that under a local Lorentz transformation
$\Lambda^{a}{}_{b}(x)$ a general spin connection $\omega^{a}{}_{b \mu}$ changes according to \cite{Kobayashi_Nomizu1996}
\begin{equation}
\omega^{a}{}_{b \mu} = \Lambda^{a}{}_{e}(x) \, \omega'^{e}{}_{d \mu} \, \Lambda_{b}{}^{d}(x) +
\Lambda^{a}{}_{e}(x) \, \partial_{\mu} \Lambda_{b}{}^{e}(x),
\label{ltsc}
\end{equation}
the spin connection \eqref{InerConn} is seen to be the connection obtained from a Lorentz transformation of a vanishing spin connection $\Aw'^e{}_{d \mu} = 0$:
\begin{equation}
\Aw^a{}_{b \mu} = \Lambda^a{}_e(x) \, \Aw'^e{}_{d \mu} \, \Lambda_b{}^d(x) +
\Lambda^a{}_e(x) \, \partial_\mu \Lambda_b{}^e(x).
\label{ltscBis}
\end{equation}
Starting from an inertial frame, in which the inertial spin connection vanishes, different classes of non--inertial frames are obtained by performing \emph{local} (point--dependent) Lorentz transformations $\Lambda^{a}{}_b(x^\mu)$. Within each class, the infinitely many frames are related through \emph{global} (point--independent) Lorentz transformations $\Lambda^{a}{}_b =$~constant.

Now, due to the orthogonality of the tetrads, transformation \eqref{LoreTrans-e} can be rewritten in the form
\begin{equation}
\Lambda^a{}_b(x) = e^a{}_\mu e'_b{}^\mu \,.
\end{equation}
Using this relation, the coefficient of anholonomy \eqref{fcab0} of the frame $e^a{}_\mu$ is found to be
\begin{equation}
f^c{}_{a b} = \Aw^c{}_{b a} - \Aw^c{}_{a b}
\label{relation00}
\end{equation}
where we have identified $\Aw^a{}_{b c} = \Aw^a{}_{b \mu} \, e_c{}^\mu$. The inverse relation is
\begin{equation}
\Aw^{a}{}_{b c} = \onehalf \left(f_{b}{}^{a}{}_{c}
+ f_{c}{}^{a}{}_{b} - f^{a}{}_{b c} \right).
\label{InerConefff}
\end{equation}
Of course, as a purely inertial connection, $\,\Aw^a{}_{b \mu}$ has vanishing curvature:
\begin{equation}
\Rw^{a}{}_{b \nu \mu} \equiv \partial_{\nu} \Aw^{a}{}_{b \mu} -
\partial_{\mu} \Aw^{a}{}_{b \nu} + \Aw^a{}_{e \nu} \, \Aw^e{}_{b \mu}
- \Aw^a{}_{e \mu} \, \Aw^e{}_{b \nu} = 0 \,.
\label{curvaDefW}
\end{equation}
For $e^a{}_\mu$ a trivial tetrad, $\Aw^a{}_{b \mu}$ has also vanishing  torsion:
\begin{equation}
\Tw^a{}_{\nu \mu} \equiv \partial_{\nu} e^{a}{}_{\mu} -
\partial_{\mu} e^{a}{}_{\nu} + \Aw^a{}_{e \nu} \, e^e{}_{\mu}
- \Aw^a{}_{e \mu} \, e^e{}_{\nu} = 0.
\label{tordefW}
\end{equation}

\subsection{Example: Equation of motion of free particles}\label{InerEquations}

In the class of inertial frames $e'^a{}_\mu$, a free particle is described by the equation of motion
\begin{equation}
\frac{d u'^a}{d\sigma} = 0,
\label{EM25}
\end{equation}
with $u'^a$ the anholonomic particle four--velocity, and
\begin{equation}
d \sigma^2 = \eta_{\mu \nu} \, dx^\mu dx^\nu
\label{MinkoInter}
\end{equation}
the quadratic Minkowski interval. Of course, since it is written in a specific class of frames, equation \eqref{EM25} is not \emph{manifestly} covariant under local Lorentz transformations. This does not mean, however, that it is not covariant. In fact, in the anholonomic frame $e^a{}_\mu$, related to $e'^a{}_\mu$ by the local Lorentz transformation \eqref{LoreTrans-e}, the equation of motion of a free particle assumes the Lorentz covariant form
\begin{equation}
\frac{d u^a}{d\sigma} + \Aw^a{}_{b \mu} \, u^b \, u^\mu = 0,
\label{anholoEM}
\end{equation}
where
\begin{equation}
u^a = \Lambda^a{}_b(x) \, u'^b
\end{equation}
is the Lorentz transformed four--velocity, with $u^\mu = u^a \, e_a{}^\mu = {d x^\mu}/{d\sigma}$ the spacetime holonomic four--velocity. In the tetrad version of special relativity, therefore, the Lorentz connection $\Aw^a{}_{b \mu}$ represents inertial effects only, and are responsible for rendering relativistic physics invariant under local Lorentz transformations.

In terms of the holonomic four--velocity $u^\rho$, the equation of motion \eqref{anholoEM} assumes the form
\begin{equation}
\frac{d u^\rho}{d\sigma} + \gammaw^\rho{}_{\nu \mu} \, u^\nu u^\mu = 0,
\label{holoEM}
\end{equation}
where
\begin{equation}
\gammaw^\rho{}_{\nu \mu} =
e_c{}^\rho \partial_\mu e^c{}_\nu + e_c{}^\rho \Aw^c{}_{b \mu} e^b{}_\nu \equiv
e_c{}^\rho \Dw_\mu e^c{}_\nu
\label{STindIneCon}
\end{equation}
is the spacetime--indexed version of the inertial spin connection $\Aw^a{}_{b \mu}$. The inverse relation is
\begin{equation}
\Aw^a{}_{b \mu} =
e^a{}_\rho \partial_\mu e_b{}^\rho + e^a{}_\rho \gammaw^\rho{}_{\nu \mu} e_b{}^\nu \equiv
e^a{}_\rho \nablaw_\mu e_b{}^\rho.
\label{STindIneCon2}
\end{equation}
Connection $\gammaw^\rho{}_{\nu \mu}$ is sometimes referred to as the \emph{Ricci coefficient of rotation} \cite{Misner_Thorne_Wheeler1973}.


\section{Basics of Teleparallel Gravity\label{secBasics}}

\subsection{Gauge structure}\label{sec:GaugeStruct}

Teleparallel gravity can be interpreted as a gauge theory for the translation group~\cite{Hayashi_Nakano1967,Aldrovandi_Pereira2013}. The reason for \emph{translations} can be understood from the gauge paradigm, of which Noether's theorem is a fundamental piece. Recall that the source of the gravitational field is energy and momentum. According to Noether's theorem, the energy--momentum current is covariantly conserved provided the source Lagrangian is invariant under spacetime translations. If gravitation is to present a gauge formulation with energy--momentum as the source, then it must be a gauge theory for the translation group.

A gauge transformation in teleparallel gravity is defined as a local translation of the tangent space coordinates,
\begin{equation}
x^a \to x^a + \varepsilon^a(x^\mu) \, ,
\end{equation}
with $\varepsilon^a(x^\mu)$ the infinitesimal transformation parameter. Under such a transformation, a general source field $\Psi = \Psi(x^a(x^\mu))$ transforms according to (see Ref.~\cite{Aldrovandi_Pereira2013}, page 42)
\begin{equation}
{\delta_\epsilon} \Psi = \varepsilon^a(x^\mu) \partial_a \Psi \, ,
\label{trafiBis}
\end{equation}
with $\partial_a$ the translation generators. For a global translation with parameter $\varepsilon^a =$ constant, the ordinary derivative $\partial_\mu \Psi$ transforms covariantly:
\begin{equation}
\delta_\epsilon(\partial_\mu \Psi) = \varepsilon^a \partial_a \big(\partial_\mu \Psi\big).
\end{equation}
For a local transformation with parameter $\varepsilon^a(x)$, however, it does not trans\-form covariantly:
\begin{equation}
\delta_\epsilon(\partial_\mu \Psi) = \varepsilon^a(x) \partial_a \big(\partial_\mu \Psi\big) +
\big(\partial_\mu \varepsilon^a(x)\big) \partial_a \Psi.
\end{equation}
In fact, the last term on the right-hand side is a spurious term, which breaks the translational gauge covariance of the transformation. Similar to all other gauge theories~\cite{Itzykson_Zuber1980}, in order to recover gauge covariance it is necessary to introduce a (in this case translational) gauge potential $B^{a}{}_{\mu}$, a 1-form assuming values in the Lie algebra of the translation group: $B_\mu = B^a{}_\mu \partial_a$.
This potential can be used to construct the gauge covariant derivative
\begin{equation}
h_\mu \Psi = \partial_\mu \Psi + B^a{}_{\mu} \, \partial_a \Psi \,
\label{TransCoVa}
\end{equation}
which holds in the class of Lorentz frames in which there are no inertial effects. In fact, provided the gauge potential transforms according to
\begin{equation}
\delta_\epsilon B^a{}_\mu = -\, \partial_\mu \varepsilon^a(x) \, ,
\label{btrans0}
\end{equation}
the derivative $h_\mu \Psi$ is easily seen to transform covariantly under gauge translations:
\begin{equation}
\delta_\epsilon (h_\mu \Psi) = \varepsilon^a(x) \partial_a (h_\mu \Psi).
\end{equation}
This is the output of the gauge construction applied to the translation group.

Owing to the soldered property of the tangent bundle,\footnote{The presence of the tetrad field provides a relationship between tangent space (internal) tensors and spacetime (external) tensors, which is what is meant by the term ``soldering."} on which teleparallel gravity is constructed, the gauge covariant derivative \eqref{TransCoVa} can be rewritten in the form
\begin{equation}
h_\mu \Psi = h^a{}_\mu \partial_a \Psi,
\label{TetraCoVa}
\end{equation}
where
\begin{equation}
h^a{}_\mu = \partial_\mu x^a + B^a{}_\mu
\label{NonTriviaTetra}
\end{equation}
is a non--trivial tetrad field. By non--trivial we mean a tetrad with $B^a{}_\mu \neq \partial_\mu \varepsilon^a$, otherwise it would be just a translational gauge transformation of the trivial tetrad $e^a{}_\mu = \partial_\mu x^a$.

Similar to any relativistic theory, the equivalent expressions valid in a general Lorentz frame can be obtained by performing a local Lorentz transformation
\begin{equation}
x^a \to \Lambda^a{}_b(x) \, x^b.
\end{equation}
Considering that the translational gauge potential $B^a{}_\mu$ is a Lorentz vector in the algebraic index; that is,
\begin{equation}
B^a{}_\mu ~\to~ \Lambda^a{}_b(x) \, B^b{}_\mu,
\label{TransLoren}
\end{equation}
it is easy to see that, in a general Lorentz frame, the translational covariant derivative \eqref{TransCoVa} assumes the form
\begin{equation}
h_\mu \Psi = \partial_\mu  \Psi + \Aw^a{}_{b \mu} x^b \, \partial_a \Psi + B^a{}_\mu  \partial_a \Psi
\label{TransCova}
\end{equation}
with $\Aw^{a}{}_{b\mu}$ the purely inertial Lorentz connection \eqref{InerConn}. The tetrad components \eqref{TetraCoVa} of the  derivative \eqref{TransCova} can then be written as
\begin{equation}
h^a{}_\mu = \partial_\mu x^a + \Aw^a{}_{b \mu} \, x^b + B^a{}_\mu \,.
\label{TeleTetrada}
\end{equation}
The first two terms on the right-hand side make up the trivial tetrad
\begin{equation}
e^a_{\ \mu} \equiv \Dw_\mu x^a= \partial_\mu x^a + \Aw^a{}_{b \mu} \, x^b \,,
\label{GenTriviTetra}
\end{equation}
which allows \eqref{TeleTetrada} to be rewritten in the form
\begin{equation}
h^a{}_\mu = \Dw_\mu x^a + B^a{}_\mu \,.
\label{TeleTetrada2}
\end{equation}
In a general class of frames, therefore, the gauge transformation of $B^a{}_\mu$ is
\begin{equation}
\delta_\epsilon B^a{}_\mu = - \, \Dw_\mu \varepsilon^a.
\label{BamGauTrans}
\end{equation}
In the class of frames in which the inertial spin connection $\Aw^{a}{}_{b\mu}$ vanishes, it assumes the form  \eqref{btrans0}.

\subsection{Translational field strength: torsion}\label{sec:FieldStrTorsion}

As in any gauge theory, the field strength of teleparallel gravity can be obtained from the commutation relation of gauge covariant derivatives. Using the translational covariant derivative [see Eq.~\eqref{TransCova}]
\begin{equation}
h_\mu \Psi = \partial_\mu  \Psi + \Aw^a{}_{b \mu} x^b \, \partial_a\Psi + B^a{}_\mu  \partial_a \Psi \, ,
\label{TransCovaBis}
\end{equation}
we obtain
\begin{equation}
[h_{\mu}, h_{\nu}] = \Tw^{a}{}_{\mu \nu}  \partial_a \, ,
\label{commu1}
\end{equation}
where
\begin{equation}
\Tw^a{}_{\mu \nu} = \partial_\mu B^a{}_\nu - \partial_\nu B^a{}_\mu +
\Aw^a{}_{b \mu} B^b{}_{\nu} - \Aw^a{}_{b \nu} B^b{}_{\mu}
\label{tfs}
\end{equation}
is the translational field strength, and $\partial_a$ stands for the translation generators. Adding the vanishing piece
\[
\Dw_\mu \big(\Dw_\nu x^a \big) - \Dw_\nu \big(\Dw_\mu x^a \big)
\equiv 0
\]
to the right--hand side of \eqref{tfs}, it becomes
\begin{equation}
\Tw^a{}_{\mu \nu} = \partial_\mu h^a{}_\nu - \partial_\nu h^a{}_\mu +
\Aw^a{}_{b \mu} h^b{}_{\nu} - \Aw^a{}_{b \nu} h^b{}_{\mu} \,.
\label{tfs2}
\end{equation}
Consequently, we see that the field strength of teleparallel gravity is just the torsion tensor. It should be noted that, using this construction, the spin connection appearing within the tetrad is the same as that appearing explicitly in the last two terms of the definition of torsion. In fact, if these connections are not the same, $\Tw^a{}_{\mu \nu}$ can no be longer be interpreted as the translational field strength. In the class of frames in which $\Aw^a{}_{b \mu}$ vanishes, torsion assumes the form
\begin{equation}
\Tw^a{}_{\mu \nu} = \partial_\mu h^a{}_\nu - \partial_\nu h^a{}_\mu
\label{tfs3}
\end{equation}
with $h^a{}_\nu$ the tetrad (\ref{NonTriviaTetra}).
In Section~\ref{secPT} we will return to discuss this point in connection to the so--called ``pure tetrad teleparallel gravity''.

Through contraction with a tetrad, the torsion tensor can be written in the form
\begin{equation}
\Tw^\rho{}_{\mu \nu} \equiv h_a{}^\rho \Tw^a{}_{\mu \nu} =
\Gammaw^{\rho}{}_{\nu \mu} - \Gammaw^{\rho}{}_{\mu \nu} \, ,
\label{WeitConn}
\end{equation}
where
\begin{equation}
\Gammaw^{\rho}{}_{\nu \mu} = h_{a}{}^{\rho} \partial_{\mu} h^{a}{}_{\nu} +
h_{a}{}^{\rho}\Aw^a{}_{b \mu} \, h^b{}_\nu \equiv
h_{a}{}^{\rho} \, \Dw_{\mu} h^{a}{}_{\nu}
\label{gecow}
\end{equation}
is the \emph{non-trivial} spacetime--indexed connection corresponding to the inertial spin connection $\Aw^a{}_{b \mu}$, also known as the Weitzenb\"ock connection. Its definition is equivalent to the identity
\begin{equation}
\partial_{\mu}h^a{}_{\nu} +
\Aw^a{}_{b \mu} \, h^b{}_\nu -
\Gammaw^{\rho}{}_{\nu \mu} \, h^a{}_{\rho} = 0.
\label{cacd1}
\end{equation}
In the class of frames in which the spin connection $\Aw^a{}_{b \mu}$ vanishes, it reduces to
\begin{equation}
\partial_{\mu}h^a{}_{\nu} -
\Gammaw^{\rho}{}_{\nu \mu} \, h^a{}_{\rho} = 0 \,.
\label{cacd0}
\end{equation}
In all other classes of frame, it assumes the general form \eqref{cacd1}.

\subsection{Gravitational coupling prescription}\label{sec:Gravicouplprescrip}

\subsubsection{Translational coupling prescription}\label{sec:translcouplprescrip}

As discussed in Section~\ref{sec:GaugeStruct}, in the absence of gravitation the Lorentz covariant derivative in a general frame is written as
\begin{equation}
e_\mu \Psi = e^a_{\ \mu} \partial_a \Psi \, ,
\end{equation}
with $e^a_{\ \mu}$ the trivial tetrad \eqref{GenTriviTetra}. In the presence of gravitation, on the other hand, it is given by
\begin{equation}
h_\mu \Psi = h^a_{\ \mu} \partial_a \Psi \, ,
\end{equation}
with $h^a_{\ \mu}$ the non-trivial tetrad \eqref{TeleTetrada}. The translational coupling prescription in a general class of frames can then be written in the form
\begin{equation}
e^a{}_\mu \partial_a \Psi ~\to~ h^a{}_\mu \partial_a \Psi \,.
\label{TransCP0}
\end{equation}
Such a coupling prescription actually amounts to the tetrad replacement
\begin{equation}
e^a{}_\mu \to h^a{}_\mu \, ,
\label{TCPrescri}
\end{equation}
which, in turn, amounts to replacing the spacetime Minkowski metric by a general Riemanian metric:
\begin{equation}
\eta_{\mu \nu} = \eta_{ab} \, e^a{}_\mu e^b{}_\nu ~\rightarrow~
g_{\mu \nu} = \eta_{ab} \, h^a{}_\mu h^b{}_\nu \,.
\label{MetricReplace}
\end{equation}
As a consequence, the spacetime intervals change according to
\begin{equation}
d\sigma^2 = \eta_{\mu \nu} dx^\mu dx^\nu ~\to~ ds^2 = g_{\mu \nu} dx^\mu dx^\nu \,.
\label{IntervalCoupling}
\end{equation}
It is important to remark that in general relativity such replacement is implicitly assumed whenever applying the gravitational coupling prescription. In teleparallel gravity, on the other hand, it emerges naturally as a consequence of the translational coupling prescription. Furthermore, in contrast to general relativity, it provides an explicit expression for the tetrad field, as given by Eq.~\eqref{TeleTetrada}.

\subsubsection{Lorentz coupling prescription}\label{sec:LoreCoupPresc}
Like any other classical field theory, since local Lorentz invariance is a fundamental symmetry of the Nature, teleparallel gravity must also be invariant under local Lorentz transformations. It should be emphasised that Lorentz invariance by itself is empty of dynamical content in the sense that any relativistic equation can be written in a Lorentz covariant form. Although not a dynamic symmetry, however, the local Lorentz invariance introduces an additional coupling prescription, which is a direct consequence of the strong equivalence principle.

The explicit form of the Lorentz gravitational coupling prescription can be obtained from the so--called {\em general covariance principle} (see Ref.~\cite{Weinberg}, Section~4.1). In its frame version~\cite{GCPframeversion}, this principle states that, by writing a special--relativistic equation in a Lorentz covariant form and then using the strong equivalence principle, it is possible to obtain its form in the presence of gravitation. The general covariance principle can be thought of as an active version of the usual (or passive) strong equivalence principle, which says that, given an equation valid in the presence of gravitation, the corresponding special--relativistic equation is recovered locally (that is, at a point or along a trajectory).

Let us start with the situation in special relativity and consider  the ordinary derivative of a general field $\Psi$.  The first step of the general covariance principle is to perform a local Lorentz transformation, such that all ordinary derivatives $\partial_\mu \Psi$ assume the Lorentz covariant form,
\begin{equation}
\partial_\mu \Psi \to {\mathcal D}_\mu \Psi = \partial_\mu \Psi + \onehalf \, e^a{}_\mu \left( f_b{}^c{}_a + f_a{}^c{}_b - f^c{}_{ba} \right) \, S_c{}^b \Psi \,,
\label{LCD0}
\end{equation}
where $f^c{}_{a b}$ is the coefficient of anholonomy \eqref{fcab0} of the trivial tetrad in the Minkowski spacetime, and $S_c{}^b$ are Lorentz generators written in the representation to which $\Psi$ belongs. The last term in the right-hand side is an inertial compensating term that enforces the Lorentz covariance of the derivative in the new Lorentz frame.

In the presence of gravitation, according to the translational coupling prescription \eqref{TCPrescri}, the trivial tetrad $e^a{}_\mu$ is replaced by the nontrivial one $h^a{}_\mu$, and the coupling prescription \eqref{LCD0} assumes the form
\begin{equation}
\partial_\mu \Psi \to {\mathcal D}_\mu \Psi = \partial_\mu \Psi + \onehalf \, h^a{}_\mu \left( f_b{}^c{}_a + f_a{}^c{}_b - f^c{}_{ba} \right) \, S_c{}^b \Psi \,
\label{LCD1}
\end{equation}
with the coefficient of anholonomy now given by
\begin{equation}
f^c{}_{a b} = h_a{}^{\mu} h_b{}^{\nu} (\partial_\nu
h^c{}_{\mu} - \partial_\mu h^c{}_{\nu} ) \,.
\end{equation}
In the specific case of teleparallel gravity, where torsion is non-vanishing, relation \eqref{relation00} assumes the form
\begin{equation}
\Aw^c{}_{ba} - \Aw^c{}_{ab} = f^c{}_{ab} + \Tw^c{}_{ab} \label{Equation_66}\,,
\end{equation}
where $\Tw^c{}_{ab}$ is the torsion of the purely inertial connection $\Aw^c{}_{ab}$. Use of this equation for three
different combinations of indices gives
\begin{equation}
\onehalf \left( f_b{}^c{}_a + f_a{}^c{}_b - f^c{}_{ba} \right) =
\Aw^c{}_{ba} - \Kw^c{}_{ba},
\label{equivaprin}
\end{equation}
with the contortion tensor
\begin{equation}
\Kw^c{}_{ba} = \onehalf \big( \Tw_b{}^c{}_a + \Tw_a{}^c{}_b - \Tw^c{}_{ba} \big).
\end{equation}
The coupling prescription in the presence of gravitation is then obtained by replacing the inertial compensating term of  \eqref{LCD1} with that given by  \eqref{equivaprin}:
\begin{equation}
\partial_\mu \Psi \to  \partial_\mu \Psi +
\onehalf \big( \Aw^c{}_{b\mu} - \Kw^c{}_{b\mu} \big) \, S_c{}^b \Psi \,,
\label{LCD3}
\end{equation}
which defines the full (translational plus Lorentz) gravitational coupling prescription in teleparallel gravity.

Now, due to the fundamental identity of the theory of Lorentz connections~\cite{Kobayashi_Nomizu1996},
\begin{equation}
\Aw^c{}_{b\mu} - \Kw^c{}_{b\mu} = \Abol^c{}_{b\mu} \,,
\label{Ricci}
\end{equation}
with $\Abol^c{}_{ba}$ the Levi--Civita spin connection, the teleparallel coupling prescription \eqref{LCD3} is found to be equivalent to the general relativity coupling prescription
\begin{equation}
\partial_\mu \Psi \to \Dbol_\mu \Psi = \partial_\mu \Psi +
\onehalf \, \Abol^c{}_{b\mu} \, S_c{}^b \Psi \,.
\label{LCD2}
\end{equation}
Since both coupling prescriptions were obtained from the general covariance principle, both are consistent with the strong equivalence principle.

\subsubsection{Separating inertial effects from gravitation}\label{sec:GravInertial}

As is well-known, the spin connection $\Abol^c{}_{b\mu}$ of general relativity includes both gravitation and inertial effects. Considering that $\Aw^c{}_{b\mu}$ represents inertial effects only, and that $\Kw^c{}_{b\mu}$ represents gravitation only, identity \eqref{Ricci} provides an elegant view of the strong equivalence principle. In fact, in a local frame in which the spin connection of general relativity vanishes, $\Abol^c{}_{b\mu} \doteq 0$, that identity reduces to
\begin{equation}
\Aw^c{}_{b\mu} \doteq \Kw^c{}_{b\mu} \,,
\label{Ricci0}
\end{equation}
from which we see that, in this local frame, inertial effects $\Aw^c{}_{b\mu}$ exactly compensates for gravitation $\Kw^c{}_{b\mu}$.

As an illustration, let us consider a free particle in Minkowski spacetime, whose equation of motion has the form
\begin{equation}
u^\mu \partial_\mu u^a = 0 \,.
\label{FreeEM}
\end{equation}
Since the four-velocity $u^a$ is a Lorentz vector, we use the vector representation of the Lorentz generators, which is given by the matrix~\cite{Ramond:1981pw}
\begin{equation}
\big(S_c{}^b \big)^a{}_d = \delta^b_d \, \delta^a_c - \eta_{cd} \, \eta^{ab} \,.
\end{equation}
In this case, the general relativity coupling prescription \eqref{LCD2} assumes the form
\begin{equation}
\partial_\mu u^a \to \Dbol_\mu u^a = \partial_\mu u^a +
\onehalf \, \Abol^c{}_{b\mu} \, (S_c{}^b \big)^a{}_d \, u^d \,.
\label{LCD2ua}
\end{equation}
When applied to the equation of motion \eqref{FreeEM} describing a free particle, it yields the geodesic equation
\begin{equation}
u^\mu \big( \partial_\mu u^a + \Abol^a{}_{b\mu} u^b \big) = 0 \,.
\label{GeodEq}
\end{equation}
The vanishing of the right-hand side means that in general relativity there is no gravitational force. In this theory, gravitation and inertial effects are described by the geometry of spacetime, and are included in the spin connection in the left-hand side of the equation.
In a similar way, applying the teleparallel coupling prescription \eqref{LCD3} to the free equation of motion \eqref{FreeEM} yields the teleparallel force equation
\begin{equation}
u^\mu \big( \partial_\mu u^a + \Aw^a{}_{b\mu} u^b \big) = \Kw^a{}_{b\mu} u^b u^\mu \,,
\end{equation}
which is, of course, equivalent to the geodesic equation \eqref{GeodEq}. In this description, however, the inertial effects, represented by the inertial spin connection $\Aw^a{}_{b\mu}$, remain geometrized in the left-hand side of the equation, whereas gravitation, represented by the contortion tensor $\Kw^a{}_{b\mu}$, plays the role of a gravitational force on the right-hand side~\cite{AndradePereira97}. This separation of gravitation and inertial effects, beautifully evinced by identity \eqref{Ricci}, is one of most prominent properties of teleparallel gravity.


\section{Lagrangian and Field Equations}\label{sec:LagrangianTeleGrav}

\subsection{Teleparallel gravity Lagrangian}

Similar to any gauge theory, the Lagrangian density of teleparallel gravity is quadratic in the torsion tensor, the field strength of the theory (we use units in which $c = 1$)
\begin{equation}
\Lw =
{\frac{h}{16 \pi G}} \left(\onefourth \; \Tw^\rho{}_{\mu\nu} \, \Tw_\rho{}^{\mu \nu} +
\onehalf  \; \Tw^\rho{}_{\mu\nu} \, \Tw^{\nu \mu}{}_\rho -
\, \Tw^\rho{}_{\mu\rho} \, \Tw^{\nu \mu}{}_\nu \right)
\label{TeleLagra}
\end{equation}
with $h = \det(h^a{}_\mu)$.

The first term corresponds to the usual Lagrangian of internal gauge theories.  The existence of the other two terms is related to the soldered character of the bundle, which allows internal and external indices to be treated on the same footing, and consequently new contractions turn out to be possible. Since torsion is a tensorial quantity, each term of this Lagrangian is invariant under both general coordinate and local Lorentz transformations. As a consequence the whole Lagrangian is also invariant, independently of the numerical value of the coefficients.

Introducing the notation $\kappa = {8 \pi G}$, we note the crucial property of the teleparallel  Lagrangian \eqref{TeleLagra} is its equivalence (up to a divergence) to the standard Einstein-Hilbert Lagrangian
\begin{equation}\label{lagequiv}
\Lw \equiv \Lbol  - \frac{1}{\kappa}\partial_\mu \Big( h \Tw^\mu\Big)
\end{equation}
where $\Tw^\mu=T^{\nu\mu}{}_{\nu}$ is the vector torsion. Due to this property teleparallel gravity is often called the \emph{teleparallel equivalent of general relativity} since the dynamical content of the field equations derived from both Lagrangians must be the same. We remark that the equivalence with general relativity is achieved only for the specific combination of numerical coefficients appearing in the Lagrangian (77). We mention in passing that those parameters can be obtained directly from the gauge paradigm, without resorting to general relativity~\cite{Aldrovandi_Pereira2013}. This is an important property in the sense that it renders teleparallel gravity a self-consistent theory.

\subsection{Teleparallel gravity field equations}

To derive the field equations of teleparallel gravity, let us consider the Lagrangian
\begin{equation}
{\mathcal L} = \Lw + {\mathcal L}_s,
\end{equation}
with ${\mathcal L}_s$ the Lagrangian of a general source field. Variation with respect to the gauge potential $B^a{}_\rho$, or equivalently, with respect to the tetrad field $h^a{}_\rho$, yields the teleparallel gravitational field equations
\begin{equation}
E_a^{\ \rho} = \kappa \, h\, {\Theta}_{a}{}^{\rho},
\label{tfe10}
\end{equation}
where on the left-hand side we have defined the Euler-Lagrange expression
\begin{equation}
E_a^{\ \rho} \equiv \kappa \frac{\delta \Lw_f}{\delta h^a_{\ \rho}} = \partial_\sigma\Big(h \sw_a{}^{\rho \sigma}\Big) -
\kappa \, h \jw_{a}{}^{\rho}
\end{equation}
and we have introduced the superpotential
 \begin{equation}
 \sw_a^{\ \rho\sigma}=\frac{1}{2}
 \left(
 \Tw^{\sigma \rho}_{\ \ \ a}
 +\Tw^{\ \rho\sigma}_{a}
 -\Tw^{\rho\sigma}_{\ \ \ a}
 \right)
 -h_a^{\ \sigma}\Tw^{\theta \rho}_{\ \ \theta}
 +h_a^{\ \rho}\Tw^{\theta \sigma}_{\ \ \theta}
 \label{sup}
 \end{equation}
and  the gauge current
\begin{equation}
\jw_{a}{}^{\rho} = \frac{1}{\kappa} h_a{}^{\lambda}\sw_c{}^{\nu\rho}\Tw^c{}_{\nu\lambda} -\frac{h_a^{\ \rho}}{h} \Lw + \frac{1}{\kappa} \omegaw^c{}_{a\sigma} \sw_c{}^{\rho\sigma},
\end{equation}
which in this case represents the Noether energy-momentum pseudo-current of gravitation~\cite{deAndrade_Guillen_Pereira2000}.

The right-hand side of the field equations \eqref{tfe10} is the matter energy-momentum tensor
\begin{equation}\label{matemten}
{\Theta}_{a}{}^{\rho} = - \frac{1}{h} \frac{\delta {\mathcal L}_s}{\delta h^a_{\ \rho}}.
\end{equation}
We note that the anti-symmetric part of $\Theta_{[\mu\nu]} = h^{a}_{\phantom{a}[\mu} g^{\vphantom{a}}_{\nu]\rho} \Theta_a^{\phantom{a}\rho}=0$ due to the invariance of the action under local Lorentz transformations~\cite{Aldrovandi_Pereira2013}.

\subsection{Alternative forms of teleparallel gravity field equations}

There are many equally valid ways to write the field equations \eqref{tfe10} that can be useful in different situations. One of them is to write the field equations using the teleparallel covariant derivative as
\begin{equation}
\Dw_\sigma\Big(h \sw_a{}^{\rho \sigma}\Big) -
\kappa \, h \Sigmaw_{a}{}^{\rho} = \kappa \, h\, {\Theta}_{a}{}^{\rho},
\label{tfcov}
\end{equation}
where we have defined the gravitational energy-momentum tensor
\begin{equation}
\Sigmaw_{a}{}^{\rho}= \jw_{a}{}^{\rho}- \frac{1}{\kappa} \omegaw^c{}_{a\sigma}\sw_c{}^{\rho\sigma}\,.
\end{equation}
The advantage of this form of the field equations is that $\Sigmaw_{a}{}^{\rho}$ is a proper tensor under both diffeomorphisms and local Lorentz transformations. Moreover, it can be shown to be trace-free:
\begin{equation}\label{key}
\Sigmaw_{\rho}{}^{\rho} \equiv h^a_{\ \rho}\Sigmaw_{a}{}^{\rho}=0,
\end{equation}
as is appropriate for a massless field.
Alternatively, the field equations \eqref{tfe10} can be rewritten in terms of spacetime-indexed  quantities as
\begin{equation}
E_\mu{}^\rho \equiv \partial_\sigma \Big(h \sw_\mu^{\ \rho\sigma} \Big)+
\kappa  h \tw_\mu^{\ \rho}=\kappa \, h\, {\Theta}_{\mu}{}^{\rho}, \label{steq}
\end{equation}
where
\begin{equation}
h \tw_\mu^{\ \rho}=   \frac{1}{\kappa} h  \Gammaw^\alpha_{\phantom{\mu}\sigma\mu}\sw_{\alpha}^{\ \sigma\rho}+\delta_\mu^{\ \rho} \Lw, \label{ptensor}
\end{equation}
is the energy-momentum pseudotensor. Due to the fact that $h \tw_\mu^{\ \rho}$ is conserved with ordinary derivative, it straightforwardly leads to spacetime conserved charges~\cite{Krssak_Pereira2015}. Moreover, the left-hand side $E_\mu{}^\rho$ of (\ref{steq}) is  symmetric, which allows an easy comparison with the field equations of general relativity in the metric formulation.

\subsection{Variations with respect to the spin connection}\label{surf}

From the point of view of the gauge approach to teleparallel gravity discussed in Section~\ref{secBasics}---in particular, from the fact that the tetrad can be written in terms of the translational potential and the spin connection, as can be seen from \eqref{TeleTetrada}---it is clear that the spin connection is not an independent variable from the tetrad. However,  if we decide to use as a starting point, not the gauge paradigm but the Lagrangian \eqref{TeleLagra} written as a function of the tetrad and the spin connection, we need to address the problem of the variation of the Lagrangian with respect to the spin connection.

As we will show shortly, it turns out that the variation with respect to the spin connection is identically satisfied for an arbitrary teleparallel spin connection, and hence there are no extra field equations that would determine it. This is consistent with our interpretation of the spin connection as representing inertial effects only, and hence should not have their own dynamics governed by extra field equations\footnote{However, as we will show later in section~\ref{secfteq}, this changes in the case of modified teleparallel theories of gravity when the spin connection becomes dynamical with non-trivial field equations---albeit not independent from the field equations for the tetrad.}. The problem of how to determine the spin connection will be discussed in detail in Section~\ref{seccon}.

There are multiple methods to  compute the variations of the Lagrangian with respect to the spin connection. Let us first introduce  the method developed in~\cite{Krssak2017}. We consider a teleparallel Lagrangian corresponding to a vanishing spin connection and a Lagrangian for an arbitrary spin connection: i.e.,
\begin{equation}\label{twotet}
\Lw(h^a_{\mu},0) \qquad \text{and} \qquad \Lw(h^a_{\mu},\Aw^a_{\ b\mu}).
\end{equation}
Then we use the equivalence of these Lagrangians with the Einstein-Hilbert Lagrangian, as given by Eq.~\eqref{lagequiv}. Since both Lagrangians \eqref{twotet} correspond to the same tetrad, the same Einstein-Hilbert Lagrangian can be associated with both of them.  Their equivalence can then be written in the form
\begin{equation}
\Lbol(h^a_{\ \mu}) \equiv \Lw(h^a_{\mu},\Aw^a_{\ b\mu}) + \partial_\mu \left[\frac{h}{\kappa} \, \Tw^{\rho\mu}{}_\rho(h^a_{\mu},\Aw^a_{\ b\mu})\right] =
\Lw(h^a_{\mu},0) + \partial_\mu \left[\frac{h}{\kappa} \, \Tw^{\rho\mu}{}_\rho(h^a_{\mu},0) \right].
\label{65}
\end{equation}
On the other hand, contracting the torsion tensor \eqref{tfs2} with $h_a^{\ \nu}$ yields
\begin{equation}
\Tw^{\rho\mu}{}_\rho(h^a_{\ \mu},\Aw^a_{\ b\mu}) = \Tw^{\rho\mu}{}_\rho(h^a_{\ \mu},0) - \Aw^\mu,
\label{ide4}
\end{equation}
where we have used the notation $\Aw^\mu=\Aw^{a}_{\ b \nu}h_a^{\ \nu}h^{b \mu}$. Plugging \eqref{ide4} into \eqref{65}, the two teleparallel Lagrangians are found to be related by~\cite{Krssak2017}:
\begin{equation}	
\Lw (h^a_{\ \mu},\Aw^a_{\ b\mu}) =
\Lw (h^a_{\ \mu},0) + \frac{1}{\kappa} \partial_\mu \big(
h \Aw^\mu \big). \label{rel}
\end{equation}
This relation shows that the inertial spin connection $\Aw^{a}_{\ b \mu}$ enters the Lagrangian as a total derivative, and hence the variation with respect to the spin connection vanishes identically
\begin{equation}\label{zerovar}
\frac{\delta \Lw}{\delta \Aw^a{}_{b \mu}} = 0 \,.
\end{equation}
Moreover, relation \eqref{rel} implies that the spin connection does not contribute to the field equations.

An alternative method to vary the action is to straightforwardly vary the action and restrict the variations to those that preserve the local flatness and the teleparallel form of the spin connection~\cite{Golovnev:2017dox}.  Since the teleparallel connection \eqref{InerConn} is entirely given by the local Lorentz transformation matrix $\Lambda^a_{\ b}$,  it is sufficient to consider only its changes under infinitesimal local Lorentz transformations
\begin{equation}\label{infvar}
\Lambda^a_{\ b}=\delta^a_{\ b}+\epsilon^a_{\ b}, \qquad \qquad \epsilon_{ab}=-\epsilon_{ba}.
\end{equation}
The variation of the spin connection is then given by
\begin{equation}\label{spinnvar}
\delta \omegaw^{ab}{}_{\mu}=\delta_\epsilon \omegaw^{ab}{}_{\mu}=\Dw_\mu \epsilon^{ab}=\partial_\mu \epsilon^{ab} + \omw^a{}_{c\mu}\epsilon^{cb} + \omw^b{}_{c\mu} \epsilon^{ac}.
\end{equation}
We can then vary the action with respect to the spin connection
\begin{equation}	
\delta_\omega  \Lw =\frac{\delta \Lw}{\delta \omw^{a b}{}_{\mu}} {\delta \omw^{a b}{}_{\mu}}= \frac{h}{2 \kappa}\sw_{ab}{}^{\mu} {\delta \omw^{a b}{}_{\mu}}=
\frac{h}{2 \kappa}\sw_{ab}{}^{\mu} \Dw_\mu \epsilon^{ab}.
\label{naivvar}
\end{equation}
Integrating by parts, taking into account that the total derivative does not contribute to the field equations, and using the antisymmetry of $\epsilon^{ab}$, we find the condition
\begin{equation}\label{tegrcond}
\Dw_\mu \left(h \sw_{[ab]}{}^{\mu}\right)=0.
\end{equation}
Using the identity~\cite{Blagojevic:2000qs}
\begin{equation}\label{blagoid}
h \sw_{[ab]}{}^\mu =\Dw_\nu \left(h h_{[a}{}^{\nu}h_{b]}{}^{\mu}\right),
\end{equation}
and the fact that the teleparallel covariant derivatives commute with each other (on account of the zero curvature), we find that the field equations for the spin connection \eqref{tegrcond} are identically satisfied.

Therefore, both methods of variation presented here lead to the same result, namely that the spin connection trivially satisfies the field equations. It is also possible to show that the constrained variational principle, where the teleparallel condition is implemented using the method of Lagrange multipliers, leads exactly to the same result~\cite{Golovnev:2017dox}.

\subsection{Solving the teleparallel gravity field equations}\label{solutions}

We discuss now a method for solving the teleparallel field equations~\cite{Krssak_Pereira2015}. To begin with, let us recall that the gravitational field equations are intricate nonlinear differential equations, for which there is not a general constructive method for obtaining solutions. One has to resort to some ad hoc procedure in which some hand work is necessary. Typically one relies on the symmetries of the solution to propose an ansatz for the metric or tetrad, thereby obtaining simpler differential equations that are easier to solve. The important point is that upon proposing some ansatz, one is most likely choosing a tetrad whose associated inertial connection $\Aw^a{}_{b \mu}$ is non-vanishing.

However, as follows from the relation \eqref{rel}, the spin connection enters the action only through a surface term and hence the field equations obtained from $\Lw (h^a_{\ \mu},\Aw^a_{\ b\mu})$ and from $\Lw (h^a_{\ \mu},0)$ are the same. This means that the field equations can be solved independently of the spin connection, which is left undetermined in the process. It should be noted that the teleparallel field equations determine only the equivalence class of tetrads with respect to the {\it local} Lorentz transformations $\Lambda^a{}_{b}(x)$. In other words, tetrads related through {\it local} Lorentz transformations
\begin{equation}
  \label{twotetrads}
  h^a{}_{\mu}\qquad \text{and} \qquad h'^{a}{}_{\mu}=\Lambda^a{}_{b}h^{b}{}_{\mu},
\end{equation}
are indistinguishable as far as the teleparallel field equations are concerned.  This is a direct consequence of the fact that the teleparallel spin connection \eqref{InerConn} is not determined by the field equations. Therefore, the field equations do not determine $\Lambda^a{}_{b}(x)$, which means that the tetrad is determined up to a local Lorentz transformation. Namely, the field equations effectively determine only the metric tensor. In the next Section we show that the same situation occurs in the tetrad formulation of general relativity.

\subsection{Comparison with the tetrad formulation of general relativity\label{secComp}}

It is useful at this point to make a comparison with general relativity. As is well-known, general relativity has both a metric and a tetrad formulation~\cite{Weinberg}. In the  metric formalism we straightforwardly calculate the Riemannian curvature using Christoffel symbols from the metric tensor. The dynamics of the metric tensor is described by the Einstein field equations,
\begin{equation}\label{EEq1}
\Rbol^\mu{}_{\nu}-\frac{1}{2} \, \delta^\mu_{\nu}\Rbol=\kappa \Theta^\mu{}_{\nu}\, ,
\end{equation}
which is essentially a set of ten field equations for the ten components of the metric tensor.

On the other hand, in the tetrad formulation the ten-components of the metric tensor are replaced by the sixteen-components of the tetrad field. The Einstein field equations in this case take the similar form
\begin{equation}\label{EEq2}
\Rbol^a{}_\nu - \frac{1}{2} \, h^a_{\ \nu} \Rbol = \kappa \Theta^a{}_\nu \, ,
\end{equation}
where $\Rbol^a{}_\nu$ is the Ricci curvature calculated directly  from the tetrad, and  is related to the spacetime-indexed Ricci curvature by
\begin{equation}{\label{curvrel1}}
\Rbol^a{}_\nu =h^a{}_\mu \Rbol^\mu{}_{\nu}.
\end{equation}
From this we can see that the tetrad form of the Einstein field equations \eqref{EEq2} is just a projection of their spacetime form \eqref{EEq1} along the tetrad components. Therefore, the dynamical content of both forms of field equations is the same and they determine only the metric tensor. This is also clear from the fact that Einstein's field equations \eqref{EEq2} are covariant under local Lorentz transformations:
\begin{equation}\label{EEq3}
{\Lambda}^c{}_{a}(x) \left( \Rbol^a{}_\nu - \frac{1}{2} \, h^a_{\ \nu} \Rbol
\right)=\kappa \, {\Lambda}^c{}_{a}(x)\, \Theta^a{}_\nu.
\end{equation}
This covariance eliminates six of the sixteen equations \eqref{EEq2}, which means that the tetrad is determined by the field equations \eqref{EEq2} only up to a local Lorentz transformation. This means that we actually determine only the metric tensor. Naturally, this is an expected result since both the metric and the tetrad formulations are just two equivalent formulations of the very same theory.


\section{Tetrad and its Associated Spin Connection \label{seccon}}

To each tetrad $h^a_{\ \mu}$ there is an associated inertial spin connection $\Aw^a_{\ b\mu}$ that describes the inertial effects present in the frame. This is clear from the fundamental form of the tetrad in teleparallel gravity, as given by Eq.~\eqref{TeleTetrada}. There is a class of frames known as {\em proper frames} characterized by a vanishing spin connection: $\{h^a{}_\mu, 0 \}$. In any other class of frames related to the proper frames by a local Lorentz transformation, the spin connection will be non-vanishing, which means that there are infinitely many pairs $\{h^a{}_\mu, \Aw^{a}{}_{b\mu} \}$. Each pair defines a different class of frames, characterized by a different inertial spin connection $\Aw^{a}{}_{b\mu}$. However, in all practical cases, it is not immediately possible to identify the spin connection of a given tetrad {$h^a_{\ \mu}$}. It is then necessary to provide a method to retrieve such a spin connection from a general tetrad.

We display here the simplest method of determining the spin connection introduced in \cite{Krssak_Pereira2015}, which is based on specifying the inertial effects present in the frame $h^a_{\ \mu}$, and then finding a spin connection that precisely compensates for those effects. It should be mentioned that this is not the only way to compute the spin connection; see, e.g., the method based on the spacetime symmetries, introduced in \cite{Hohmann:2019nat}.

\subsection{Determining the inertial spin connection}
We begin by defining a ``reference tetrad", $\tref^{\;a}{}_{\mu},$ as a tetrad in which gravity is switched-off. It is, of course, a trivial tetrad in the sense that it relates two Minkowski metrics written in different coordinate systems.  This can be done by setting the gravitational constant $G$ equal to zero:
\begin{equation}
\tref^{\;a}{}_{\mu}\equiv\left. h^a_{\ \mu}\right|_{G\rightarrow 0}. \label{reftet}
\end{equation}
In such a tetrad, the gravitational potential $B^a_{\ \mu}$ does not appear and the reference tetrad can be written formally as
\begin{equation}
\tref^{\;a}{}_{\mu}=\partial_\mu x^a + \Aw^a_{\ b\mu} x^b. \label{inerdecomp}
\end{equation}
Furthermore, considering that this tetrad represents a trivial frame (see Sec.~\ref{sec:frames}), the torsion tensor of the spin connection $\Aw^a_{\ b\mu}$ vanishes identically:
\begin{equation}
\Tw^a_{\ \mu\nu}(\tref^{\;a}{}_{\mu}, \Aw^a_{\ b\mu})= 0. \label{torzero}
\end{equation}

The coefficients of anholonomy $f^c{}_{a b}$ of the general tetrad $h^a_{\ \mu}$, according to Eq.~\eqref{fcab0}, are given by
\begin{equation}
f^c{}_{a b} = h_a{}^{\mu} h_b{}^{\nu} (\partial_\nu
h^c{}_{\mu} - \partial_\mu h^c{}_{\nu} ).
\label{fcab}
\end{equation}
Using Eq.~\eqref{Equation_66}, where the torsion is written in terms of $f^c{}_{a b}$ as
\begin{equation}
\Tw^a{}_{bc} = - f^a{}_{bc} + (\Aw^a{}_{cb} - \Aw^a{}_{bc}) \, ,
\label{AnhoTor}
\end{equation}
we find that the condition \eqref{torzero} for the reference tetrad assumes the form
\begin{equation}
\Tw^a{}_{bc} (\tref^{\;a}{}_{ \mu},\Aw^a_{\ b\mu}) = \Aw^a{}_{cb} - \Aw^a{}_{bc} - f^a{}_{bc}(\tref) =  0,
\label{toree}
\end{equation}
with $f^a{}_{bc}(h_{(\rm r)})$ the coefficients of anholonomy of the reference tetrad $h_{(\rm r)}^{\;a}{}_{\mu}$. Using \eqref{toree} for three different combination of indices, we can solve for the spin connection \cite{Krssak_Pereira2015}:
\begin{equation}
\Aw^a{}_{b\mu} = \onehalf h_{(\rm r)}^{\;c}{}_{\mu} \Big[f_b{}^a{}_c(h_{(\rm r)}) + f_c{}^a{}_b(h_{(\rm r)}) - f^a{}_{bc}(h_{(\rm r)}) \Big].
\label{regconsolution}
\end{equation}
This is the inertial spin connection naturally associated to the reference tetrad $h_{(\rm r)}^{\;a}{}_{\mu}$. Since the reference tetrad $h_{(\rm r)}^{\;a}{}_{\mu}$ and the original tetrad $h^a_{\ \mu}$ differ only by their gravitational content---the inertial content of both tetrads are the same---the spin connection \eqref{regconsolution} is the inertial spin connection naturally associated to the original tetrad $h^a{}_\mu$ as well. Notice, in addition, that the expression for the teleparallel spin connection \eqref{regconsolution} coincides with the Levi-Civita spin connection for the reference tetrad, and hence we can write
\begin{equation}
\Aw^a{}_{b\mu} (h^a_{\ \mu}) =\Abol^a{}_{b\mu} (h_{(\rm r)}^{\;a}{}_{\mu}).
\label{regconsolution2}
\end{equation}
We would like to stress that the Levi-Civita connection is calculated for the reference tetrad corresponding to the Minkowski spacetime and consequently it is guaranteed to have a vanishing curvature; it is  hence  in the class of teleparallel connections \eqref{InerConn}.

The crucial point is that the torsion tensor \begin{equation}
\Tw^a_{\ \mu \nu} (h^a_{\ \mu},\omega^a_{\ b\mu}) \label{gtor}
\end{equation}
constructed from the ``full'' tetrad and the spin connection
represents purely gravitational torsion in the sense that the spurious contribution from the inertial effects are removed.

\subsection{The regularizing role of the inertial spin connection}

Let us begin by considering an action for the reference tetrad \eqref{reftet}, which represents only inertial effects. If we naively associate a vanishing spin connection to the reference tetrad $\tref^{\;a}{}_{\mu}$, the gravitational action assumes the form
\begin{equation}
\Sw(\tref^{\;a}{}_\mu, 0 )=\int_\mathcal{M}\Lw(\tref^{\;a}{}_\mu, 0 ).
\label{reftetact}
\end{equation}
In general this action does not vanish, and is even typically divergent. The reason for this is that it is an action for inertial effects, which in general do not vanish at infinity \cite{Landau_Lifshitz1975}. If, instead of a vanishing spin connection, we choose the appropriate inertial spin connection \eqref{regconsolution}, then from Eq.~\eqref{torzero} we have
\begin{equation}
\Sw(\tref^{\;a}{}_\mu, \Aw^a_{\ b\mu} )=0. \label{srefac}
\end{equation}
We now see that the role of spin connection $\Aw^a_{\ b\mu} $ is to remove all inertial effects of the action, in such a way that it now vanishes---as it should because it represents only inertial effects.

From the point of view of inertial effects, the full and reference tetrads are equivalent in the sense that their inertial content are the same. This consequently means that the spin connection associated with the full tetrad (or reference tetrad) is able to remove the inertial contributions, not only from the inertial action, but from the full action as well. This yields an action that represents gravitational effects only. Considering that the inertial effects are responsible for causing the divergences, the purely gravitational action with the appropriate spin connection,
\begin{equation}
\Sw_{\rm ren}=\int_\mathcal{M} \Lw(h^a{}_{\mu},\Aw^a{}_{b\mu}), \label{actren}
\end{equation}
will always be finite for any solution of the gravitational field equations. It
can consequently be viewed as a \emph{renormalized action} \cite{Krssak_Pereira2015}.

We note that it is possible to achieve the same results in a simpler way. In fact, relation \eqref{rel} shows that the divergences are removed from the action by adding an appropriate surface term to the action, which is analogous to the process of holographic renormalization. However, in teleparallel gravity it can be interpreted as the removal of the spurious inertial effects from the theory.  Of course, once the spurious inertial contribution to the Lagrangian is removed, all quantities computed using this Lagrangian, such as for example energy and momentum, will also be finite \cite{Lucas_Obukhov_Pereira2009}.

Furthermore, there is an important difference in relation to other renormalization methods: the inertial effects in teleparallel gravity are removed locally at each point of spacetime and not from the whole integral, as it happens to be the case in other formalisms. It is then possible to define at each point the energy and momentum densities of the gravitational field \cite{Krssak_Pereira2015}. However, it should be kept in mind that this can be achieved only with the help of the reference tetrad \cite{Krssak:2017nlv}.


\section{Example: The Spherically Symmetric Vacuum Solution\label{secExample}}

\subsection{Setting up the problem}

Let us now illustrate with an explicit example the whole process of solving the field equations and determining the spin connection within the framework of teleparallel gravity. The example chosen is the static spherically symmetric problem with the well-known Schwarzschild solution. The starting point of obtaining the spherically symmetric solution in teleparallel gravity is the same as in general relativity: based on the symmetry of the problem, one proposes the ansatz metric
\begin{equation}
	ds^2=A(r)^2 dt^2-B(r)^2 dr^2-r^2d\theta^2-r^2\sin^2\theta d\phi^2 \,,
	\label{schwmet}
\end{equation}
where $A=A(r)$ and $B=B(r)$ are arbitrary functions to be determined from the field equations.

As we have already discussed, there are infinitely many tetrads corresponding to the metric ansatz \eqref{schwmet}.  We consider here two tetrads: the diagonal tetrad
\begin{equation}
h^a_{\ \mu}=\text{diag}\left( A,
B,r, r\sin \theta \right) \,
\label{tetdiag}
\end{equation}
and the off-diagonal tetrad
\begin{equation}\label{tetprop}
\tilde{h}^a_{\ \mu}=
\left( \begin{array}{cccc}
A & 0 & 0 & 0\\
0 &B \cos\phi\sin\theta & r\cos\phi\cos\theta & -r \sin\phi\sin\theta \\
0 & -B\cos\theta & r \sin\theta & 0 \\
0 & B\sin\phi\sin\theta & r\sin\phi\cos\theta & r\cos\phi\sin\theta
\end{array}
\right)\,.
\end{equation}
These two tetrads are related by
\begin{equation}\label{tetrelation}
\tilde{h}^a{}_{\mu}=\tilde{\Lambda}^a{}_{b}(x) \, h^b_{\ \mu},
\end{equation}
where $\tilde{\Lambda}^a{}_{b}(x)$ is the local Lorentz transformation
\begin{equation}
\tilde{\Lambda}^a_{\ b}=
\left( \begin{array}{cccc}
1 & 0 & 0 & 0\\
0 & \cos\phi\sin\theta & \cos\phi\cos\theta & -\sin\phi \\
0 & -\cos\theta & \sin\theta & 0 \\
0 & \sin\phi\sin\theta & \sin\phi\cos\theta & \cos\phi
\end{array}
\right).
\label{specLorentz}
\end{equation}
Obviously both tetrads  represent the same metric \eqref{schwmet} because, as is well known, the metric is invariant under local Lorentz transformations.

\subsection{Solving the field equations}
We choose to solve the field equations in the spacetime form \eqref{steq}, and we assume a zero spin connection initially. We proceed then to obtain all geometrical objects for both the diagonal and the non-diagonal tetrads.

\subsubsection{Using the diagonal tetrad \texorpdfstring{$h^a_{\ \mu}$}{diagonal h}}
For the case of the diagonal ansatz, the non-vanishing components of the superpotential $\sw_\mu{}^{\rho\sigma}=\sw_\mu{}^{\rho\sigma}(h^a_{\ \mu}, 0)$ are found to be
\begin{equation}\label{SP1}
\sw_t{}^{tr}=-\frac{2}{r B^2},\qquad
\sw_t{}^{t\theta}=\sw_r{}^{r\theta}=-\frac{\cot\theta}{r^2}, \qquad
\sw_\theta{}^{r\theta}=\sw_\phi{}^{r\phi}=\frac{1}{rB^2} + \frac{A'}{AB^2},
\end{equation}
where we do not explicitly display the antisymmetric components $\sw_\mu{}^{\rho\sigma}=-\sw_\mu{}^{\sigma\rho}$. The non-vanishing components of the energy-momentum pseudotensor $\tw_\mu{}^{\rho}=\tw_\mu{}^{\rho}(h^a_{\ \mu}, 0)$ are
\begin{align}
\tw_t{}^t        &=-\tw_r{}^r=\tw_\theta{}^\theta=\tw_\phi{}^\phi=\frac{1}{\kappa}\frac{A+2rA'}{r^2AB^2},\nonumber\\
\tw_r{}^\theta   &=-\frac{1}{\kappa}\frac{B A'+A B'}{r^2 A B}\cot\theta,\nonumber\\
\tw_\theta{}^r   &=-\frac{1}{\kappa}\frac{A+r A'}{ r A B^2}\cot\theta\,.
\label{emptdiag}
\end{align}
Combining them, the nontrivial components of the field equations \eqref{steq} are found to be
\begin{align}
E_t{}^t            &=\left(\frac{-B+B^3+2rB'}{r^2B^3}\right)h,\label{schweq1}\\
E_r{}^r            &=\left(\frac{-A+AB^2-2rA'}{r^2AB^2}\right)h,\\
E_\theta{}^\theta  &=E_\phi{}^\phi=\left(\frac{B'(A+rA')-B(A'+rA'')}{rAB^3}\right)h. \label{schweq3}
\end{align}
One verifies that the third equation is not an independent equation. Using then the first two equations, we find the same solution as in general relativity; that is,
\begin{equation}
A(r)=\frac{1}{B(r)}=\sqrt{1 - \frac{c_1}{r}}.
\label{c1sol}
\end{equation}
Matching the solution to the Newtonian limit, the integration constant $c_1$, as in general relativity, is found to be $c_1=2GM$.

\subsubsection{Using the off-diagonal tetrad  \texorpdfstring{$\tilde{h}^a_{\ \mu}$}{Off-diagonal-h}}
It is an interesting exercise to derive the field equations for the off-diagonal ansatz $\tilde{h}^a_{\ \mu}$, which explicitly illustrates that the field equations determine the tetrad up to a local Lorentz transformation. In this case the non-vanishing components of the superpotential  $\sw_\mu{}^{\rho\sigma}=\sw_\mu{}^{\rho\sigma}(\tilde{h}^a_{\ \mu}, 0)$ are:
\begin{equation}\label{SPtil}
\sw_t{}^{tr}=\frac{2(B-1)}{r B^2},\qquad
\sw_\theta{}^{r\theta}=\sw_\phi{}^{r\phi}=\frac{-A(B-1)+rA'}{rAB^2}.
\end{equation}
Similarly, the non-vanishing components of the energy-momentum pseudotensor $\tw_\mu{}^{\rho}=\tw_\mu{}^{\rho}(\tilde{h}^a_{\ \mu}, 0)$ are:
\begin{align}
\tw_t{}^t           &=\frac{1}{\kappa}\frac{(B-1)(A(B-1)-2rA')}{r^2AB^2}, &
\tw_r{}^r           &=\frac{1}{\kappa}\frac{A(B^2-1)-2rA'}{r^2AB^2} \nonumber\\
\tw_\theta{}^\theta &=\tw_\phi{}^\phi=-\frac{1}{\kappa}\frac{A(B-1)+r(B-2)A'}{r^2AB^2},&
\tw_\theta{}^r      &=\frac{1}{\kappa}\frac{A(B-1)-rA'}{rAB^2}\cot\theta,\quad\label{emptschwtil}
\end{align}
Note that both the superpotential and the energy-momentum pseudotensors for the diagonal tetrad \eqref{tetdiag} and for the off-diagonal tetrad \eqref{tetprop} are completely different, due to the fact that when we assumed a vanishing spin connection for both tetrads both quantities $\sw_\mu{}^{\rho\sigma}$ and $\tw_\mu{}^{\rho}$ become non-tensorial in nature. In the next section we are going to compute the spin connection associated to each tetrad, and then show how the corresponding tensorial quantities $\sw_\mu{}^{\rho\sigma}$ and $\tw_\mu{}^{\rho}$ will indeed transform properly.

\subsubsection{Determining the associated spin connection}\label{schw}

We use the method introduced in Section~\ref{seccon} to find the components of the spin connection associated with the tetrads \eqref{tetdiag} and \eqref{tetprop}. The starting point is to define the reference tetrad, which for the diagonal tetrad \eqref{tetdiag} is
\begin{equation}
h_{(\rm r)}^{\;a}{}_{\mu} \equiv
\left. h^a_{\ \mu}\right|_{G \to 0}=
\text{diag}\left( 1, 1, r, r\sin \theta \right).
\end{equation}
Using \eqref{regconsolution}, we find that the non-vanishing components of the spin connection are
\begin{equation}
\Aw^{\hat{1}}_{\ \hat{2}\theta}=-\Aw^{\hat{2}}_{\ \hat{1}\theta}=-1, \quad
\Aw^{\hat{1}}_{\ \hat{3}\phi}=-\Aw^{\hat{3}}_{\ \hat{1}\phi}=-\sin\theta, \quad
\Aw^{\hat{2}}_{\ \hat{3}\phi}=-\Aw^{\hat{3}}_{\ \hat{2}\phi}=-\cos\theta. \label{spschw}
\end{equation}
These components represent the inertial effects present in the diagonal tetrad \eqref{tetdiag}.
We can analogously define the reference tetrad for the off-diagonal tetrad \eqref{tetprop} as
\begin{equation}\label{tetpropref}
\tilde{h}_{(\rm r)}^{\;a}{}_{\mu} \equiv
\left. \tilde{h}^a_{\ \mu}\right|_{G \to 0}=
\left( \begin{array}{cccc}
1 & 0 & 0 & 0\\
0 & \cos\phi\sin\theta & r\cos\phi\cos\theta & -r \sin\phi\sin\theta \\
0 & -\cos\theta & r \sin\theta & 0 \\
0 & \sin\phi\sin\theta & r\sin\phi\cos\theta & r\cos\phi\sin\theta
\end{array}
\right).
\end{equation}
Using \eqref{regconsolution}, we find that the corresponding connection vanishes
\begin{equation}\label{key1}
{\tilde{\Aw}}{}^{a}{}_{b\mu}=0.
\end{equation}
This means that the off-diagonal tetrad \eqref{tetprop} is indeed the proper tetrad, and as such it represents gravitation only. On the other hand, the diagonal tetrad \eqref{tetdiag} is not proper since the associated inertial effects, represented by the spin connection \eqref{spschw}, do not vanish.

We can now check that both tetrads with their associated spin connections do lead to the same prediction for all geometric quantities, that is,
\begin{equation}\label{key2}
\sw_\mu{}^{\rho\sigma}(\tilde{h}^a_{\ \mu}, 0)=\sw_\mu{}^{\rho\sigma}(h^a_{\ \mu}, \Aw^{a}_{\ b\mu}),
\qquad
\tw_\mu{}^{\rho}(\tilde{h}^a_{\ \mu}, 0)=\tw_\mu{}^{\rho}(h^a_{\ \mu}, \Aw^{a}_{\ b\mu}).
\end{equation}
Such quantities transform covariantly under both diffeomorphisms and local Lorentz transformations. Furthermore, they now represent only gravitation, to the exclusion of the spurious inertial effects.

\subsection{Regularization of the action}
To see how the spurious inertial effects come into play and why we need to compute the associated spin connection, we consider the action and the corresponding conserved charges~\cite{Krssak_Pereira2015}. If we compute the action using the diagonal tetrad \eqref{tetdiag} and vanishing spin connection, we find that
\begin{equation}
\actionw(h^a_{\ \mu},0)=\frac{1}{\kappa}\int_\mathcal{M}d^4 x \sin\theta, \label{schwactnon}
\end{equation}
which is obviously a divergent quantity and consequently leads to divergent conserved charges. This is precisely due to the fact that the torsion scalar $\Lw(h^a_{\ \mu},0)$---and hence the action \eqref{schwactnon}---in addition to gravitation, also includes the spurious inertial effects, which means it does not vanish at infinity. The overall integral then leads to a divergent action.
However, if we remove the inertial effects by taking into account the associated spin connection \eqref{spschw}, or equivalently use the proper tetrad \eqref{tetprop}, then we find the well-behaved renormalized action
\begin{equation}
\actionw_{\rm ren}(h^a{}_{\mu},\Aw^a_{\ b\mu})=\actionw_{\rm ren}(\tilde{h}^a{}_{\mu},0)=\frac{2}{\kappa}
\int_\mathcal{M}d^4 x \left[1+\frac{(GM-r)}{rA} \right]
\sin\theta = \int dt M. \label{schwrenact}
\end{equation}
The conserved charges obtained from this action are finite, and represent the correct physical conserved charges.


\section{Some Remarks on the Pure Tetrad Teleparallel Gravity\label{secPT}}

As already discussed, to any tetrad $h^a{}_\mu$ there is associated a specific inertial spin connection $\Aw^a{}_{b\mu}$ that represents the inertial effects present in that frame:
\begin{equation}
h^a{}_\mu = h^a{}_\mu(\Aw^a{}_{b\mu}) \,.
\label{tetradspin}
\end{equation}
On the other hand, torsion is defined as the covariant derivative of the tetrad:
\begin{equation}
\Tw^a{}_{\mu \nu} = \partial_\mu h^a{}_\nu - \partial_\nu h^a{}_\mu +
\Aw^a{}_{b \mu} h^b{}_{\nu} - \Aw^a{}_{b \nu} h^b{}_{\mu} \,.
\label{tfs2bis}
\end{equation}
Note that the  spin connection is essential for torsion \eqref{tfs2bis} to be a tensorial quantity; that is, an object that transforms covariantly under both local Lorentz and general coordinate transformations. As a consequence, the action of teleparallel gravity, which similarly to the action of any gauge theory is quadratic in the field strength (in this case, torsion), will be \emph{invariant} under both local Lorentz and general coordinate transformations. Of course, the corresponding field equations will transform \emph{covariantly} under those transformations.

The crucial observation of teleparallel gravity is that the spin connection associated with the tetrad \eqref{tetradspin} and the one used in the torsion \eqref{tfs2bis} are the very same spin connection. This is particularly clear within the approach to teleparallel gravity as a gauge theory for the translation group. As can be seen in  Section~\ref{sec:FieldStrTorsion}, the  torsion tensor is explicitly constructed in a such way that  the spin connection $\Aw^a{}_{b \mu}$ appearing in the covariant derivative is the same spin connection associated to the given tetrad $h^a{}_\mu$  according to \eqref{TeleTetrada}.

Nevertheless, it is important to note that in practical calculations, we typically start with some ansatz tetrad that suits the symmetry of the problem under consideration, and for which we do not know \textit{a priori} its associated spin connection. Due to the peculiar structure of the  teleparallel action discussed in section~\ref{surf}, we can first solve the problem for the tetrad and then determine the spin connection according to the method discussed in Section~\ref{seccon}.  Only then do the physical quantities computed from the torsion tensor, such as for example the action of the gravitational field or the gravitational conserved charges, give the correct, finite, physically relevant results. Therefore, the requirement of the finiteness of the action and conserved charges motivates the necessity to associate to each tetrad a spin connection according to the method discussed in Section~\ref{seccon}.

Now, there is in the literature a different approach to teleparallel gravity, known as \emph{pure tetrad teleparallel gravity} (see Ref.~\cite{Maluf2013} for a review). Its name stems from the fact that torsion is defined not as the covariant derivative of the tetrad~\cite{Kobayashi_Nomizu1996}, but instead as an ordinary derivative,
\begin{equation}\label{zerotor}
\Tw_{\mbox{\tiny$0$}}^{\,a}{}_{\mu \nu} = \partial_\mu h^a{}_\nu - \partial_\nu h^a{}_\mu \,,
\end{equation}
with the subscript {\it zero} denoting the ``torsion'' of the pure tetrad theory. In other words, the spin connection appearing explicitly in the torsion definition \eqref{tfs2bis} is always assumed to vanish, although to each tetrad there is associated a (generally) non-vanishing spin connection \eqref{tetradspin}. The pure tetrad formulation ignores the this fact and treats the tetrad and the spin connection as genuinely independent variables. Due to the pure gauge form  \eqref{InerConn}, the  teleparallel spin connection can be transformed to zero independently of transformations of the tetrad  and hence the vanishing spin connection can be associated with each tetrad.

There are multiple problems with this approach. First, even though \eqref{zerotor} is mathematically and physically meaningful, it is not the \emph{torsion} tensor since it does not transform as a tensor under local Lorentz transformations. As a matter of fact, it is minus the \emph{coefficient of anholonomy} $f^c{}_{a b}$ of the frame $h^a{}_\mu$, whose components, according to Eq.~\eqref{fcab}, are given by
\begin{equation}
\Tw_{\mbox{\tiny$0$}}^{\,c}{}_{a b} \equiv - f^c{}_{a b} = h_a{}^{\mu} h_b{}^{\nu} \left( \partial_\mu h^c{}_{\nu} - \partial_\nu h^c{}_{\mu} \right).
\label{fcab2}
\end{equation}
Only in the class of frames in which the inertial spin connection $\Aw^a{}_{b\mu}$ vanishes will the coefficient of anholonomy coincide with the torsion tensor. In all other classes of frames, they will not coincide. 

In spite of this problem, it is still possible to use the pure tetrad teleparallel gravity for some specific purposes. The Lagrangian of the pure tetrad teleparallel gravity is obtained from the Lagrangian~\eqref{TeleLagra} by setting the spin connection to zero and  replacing torsion by the coefficient of anholonomy 
\begin{equation}
\Lw =
{\frac{h}{16 \pi G}} \left({\textstyle{\frac{1}{4}}} \, f^a{}_{bc} f_a{}^{bc} +
{\textstyle{\frac{1}{2}}} \, f^a{}_{bc} \, f^{c b}{}_a -
\, f^a{}_{ba} \, f^{c b}{}_c \right).
\label{Cho01}
\end{equation}
Despite the fact that  $f^a{}_{bc}$ is not a tensor, the field equations derived from this Lagrangian are the same as those derived from  \eqref{TeleLagra} and it then follows, as discussed in Section~\ref{surf}, that the spin connection does not contribute to the field equations. Therefore, as far as we are interested in the solutions of the field equations, both approaches, our invariant approach and the pure tetrad one, lead to the same result.

Second, the Lagrangian \eqref{Cho01} is not invariant with respect to local Lorentz transformations, but only quasi-invariant; i.e., changes by the surface term \cite{Cho1976}. Nevertheless, these contributions through the surface term do play an important role when the total value of the action is considered and in derivation of the conserved charges. It is then found that only for certain preferred class of the frames do we obtain finite, physically relevant results. 

Third, local Lorentz symmmetry is a fundamental symmetry of general relativity. Considering that the invariant formulation of teleparallel gravity discussed here preserves this symmetry, it is the only theory that can be interpreted as the  \textit{teleparallel equivalent of general relativity}.

In the literature,  properties of this pure tetrad formulation are often erroneously attributed to teleparallel gravity itself. For example, it is quite common to find statements that in teleparallel gravity torsion is not a tensor, or that the theory is not  invariant under local Lorentz transformations, or still that there are preferred frames. Obviously all of these statements apply to the pure tetrad formulation, but not to the invariant formulation of the theory.

This confusion can be easily understood from a modern perspective where the pure tetrad teleparallel gravity can be viewed as teleparallel gravity written in the specific class of proper frames, in which the spin connection vanishes. After fixing the class of frames, the theory is no longer manifestly invariant under local Lorentz transformations, though it remains invariant under global Lorentz transformations.  The whole discussion of local Lorentz invariance in pure tetrad formulation can then be viewed as rather misguided and not an indicator of any problem of teleparallel gravity. The analogous situation in electromagnetism would be to discuss gauge invariance after fixing a specific class of gauge (the Coulomb gauge, for example), which obviously does not make sense.



\section{Modification of Teleparallel Gravity: \texorpdfstring{$f(T)$}{f(T)} Gravity\label{secfT}}
The discovery of the accelerated expansion  of the Universe has motivated the study of various extensions of general relativity. A very popular extension is the so-called $f(R)$ gravity where the Lagrangian is taken to be a function of the Ricci scalar. This relatively simple model has a number of interesting features and rich phenomenology~\cite{Sotiriou:2008rp}.

In  a similar fashion to $f(R)$ gravity,  Ferraro and Fiorini~\cite{Ferraro:2006jd,Bengochea:2008gz,Ferraro:2008ey,Linder:2010py} have proposed the $f(T)$ gravity model, where the Lagrangian is given by
\begin{equation}\label{ftaction}
\Lw_{f}=\frac{h}{2\kappa} f(\Tw),
\end{equation}
where $\Tw$ is the so-called torsion scalar representing the same quadratic torsion pieces appearing in the Lagrangian of teleparallel gravity \eqref{TeleLagra}; i.e.,\footnote{Note that here we follow the notation used in teleparallel gravity where the superpotential is defined as \eqref{sup}, while in $f(T)$ gravity the superpotential is usually defined as one half of this quantity. To bridge this difference we include this one half in the definition of the torsion scalar.}
\begin{equation}
\Tw=\frac{1}{2} \sw_a{}^{\mu\nu} \Tw^a{}_{\mu\nu} =\frac{1}{4}\Tw^\rho{}_{\mu\nu} \, \Tw_\rho{}^{\mu \nu} +
\onehalf  \; \Tw^\rho{}_{\mu\nu} \, \Tw^{\nu \mu}{}_\rho -
\, \Tw^\rho{}_{\mu\rho} \, \Tw^{\nu \mu}{}_\nu.
\label{Tscalar}
\end{equation}
Following the equivalence between the teleparallel and Einstein-Hilbert Lagrangians \eqref{lagequiv}, we can obtain the relation between the torsion and curvature scalars
\begin{align}
\Rbol = -\Tw + \Bw\,,
\label{c1}
\end{align}
where
\begin{equation}\label{boundary}
\Bw=-\frac{2}{h}\partial_\mu (h \Tw^{\mu}),
\end{equation}
is the so-called boundary term.

The boundary term does not contribute to the field equations in the case of teleparallel gravity. However, modified gravity models based on the idea of replacing the actions linear in $\Rbol$ or linear in $\Tw$ with arbitrary non-linear functions, $f(R)$ and $f(T)$ respectively, are no longer equivalent. This simply follows from the fact that an arbitrary function of a boundary term is, in general, no longer a boundary term. As we discuss later in Section~\ref{secModHD}, it is possible to relate $f(R)$ gravity and $f(T)$ gravity if we consider teleparallel gravity theories with higher derivative terms in torsion.

\subsection{Field equations and variations of the action in \texorpdfstring{$f(T)$}{f(T)} gravity\label{secfteq}}

The Lagrangian of $f(T)$ gravity is a function of both the tetrad and the spin connection, and hence we should consider variations with respect to both variables. We remind the reader about the situation in the ordinary teleparallel gravity discussed in Section~\ref{sec:LagrangianTeleGrav}, where the variation with respect to the spin connection turned out to be trivial \eqref{zerovar}. This led us to the conclusion that the field equations do not determine the spin connection and, instead, it needed to be calculated using the reference spacetime as discussed in Section~\ref{seccon}.

The situation is rather different when we consider modified teleparallel gravity models such as $f(T)$ gravity. As we will show shortly, we find that the variations with respect to both the tetrad and the spin connection are non-trivial.  However, the two variations are closely related since the variation with respect to the spin connection leads to the identical field equations as the antisymmetric part of the field equations for the tetrad.

Let us consider the $f(T)$ Lagrangian \eqref{ftaction} and a general source field
\begin{equation}
{\mathcal L} = \Lw_f + {\mathcal L}_s,
\end{equation}
and vary the action with respect to the tetrad.  This leads to the field equations for the tetrad
\begin{equation}\label{ftequation}
E_a^{\ \mu}=\kappa h \Theta_a^{\ \mu},
\end{equation}
where the Euler-Lagrange expression on the left-hand side (see~\cite{Krssak_Saridakis2015} for details) is given by
\begin{equation}\label{fTELexp}
E_a^{\ \mu}\equiv \kappa \frac{\delta \Lw_f}{\delta h^a_{\ \mu}} = f_T \partial_{\nu}\left( h \sw_a^{\ \mu\nu} \right)
+
h\left( f_{TT} \sw_a^{\ \mu\nu} \partial_{\nu} \Tw
-
f_T \Tw^b_{\ \nu a }\sw_b^{\ \nu\mu}
+
f_T \omega^b_{\ a\nu}\sw_b^{\ \nu\mu}
+
\frac{1}{2}f h_a^{\ \mu}\right),
\end{equation}
where $f_T$ and $f_{TT}$ denotes first and second order derivatives of the $f$-function with respect to the torsion scalar.

In order to analyze the symmetric and antisymmetric part of the field equations, it is useful to define the fully Lorentz-indexed Euler-Lagrange expression
\begin{equation}
E_{ab}= \eta_{bc} h^c_{\ \mu}E_a^{\ \mu},
\end{equation}
which can be explicitly written as
\begin{equation}\label{ftequationG}
E_{ab}=h\left(f_{TT} \sw_{ab}{}^{\nu} \partial_{\nu} \Tw+f_T \Gbol_{ab} +\frac{1}{2} \eta_{ab}(f-\Tw\, f_T)\right),
\end{equation}
where $\Gbol_{ab}$ is the symmetric Einstein tensor of the Levi-Civita connection calculated from the tetrad only. In this form it is straightforward to see that the last two terms are symmetric and hence the antisymmetric part of the Euler-Lagrange expression is given by
\begin{equation}
E_{[ab]}=h f_{TT}\sw_{[ab]}{}^\nu \partial_\nu \Tw.
\end{equation}
Given that the energy-momentum tensor has vanishing anti-symmetric part, $\Theta_{[ab]}=0$, the antisymmetric part of the field equations is
\begin{equation}\label{fteqas}
f_{TT}\sw_{[ab]}{}^\nu \partial_\nu \Tw =0.
\end{equation}
We can now consider the variation with respect to the spin connection and show that it leads to the same equations as \eqref{fteqas}. We follow here the method introduced in~\cite{Krssak:2017nlv}, but it is possible to derive the same result using alternative methods~\cite{Golovnev:2017dox,Hohmann:2017duq}.

The relation \eqref{rel} for the torsion scalar can be re-expressed as
\begin{equation}	
\Tw (h^a_{\ \mu},\omega^a_{\ b\mu})=
\Tw (h^a_{\ \mu},0) + \frac{2}{h}\partial_\mu \left(h\, \Aw^a{}_{b\nu} h_a{}^\nu h_c{}^{\mu} \eta^{bc}\right).  \label{rel2}
\end{equation}
We then find that the variation of ${\delta_\omega \Lw_f}$ with respect to the spin connection is given by
\begin{equation}\label{varspin2}
{\delta_\omega \Lw_f} =\frac{1}{2\kappa} h\,h_a^{\ \mu}h_b^{\ \nu}( \partial_\nu f_T) \delta \omegaw^{ab}{}_{\mu}.
\end{equation}
We can then use \eqref{spinnvar} and integrate by parts to find the condition
\begin{equation}\label{fteqspin}
f_{TT}\,\partial_\nu \Tw \, \Dw_\mu \left(h h_{[a}{}^{\mu}h_{b]}{}^{\nu}\right)=0,
\end{equation}
where we have used $\partial_\nu f_T=f_{TT}\partial_\nu \Tw$. Using the identity  \eqref{blagoid}, we find that the field equations for the spin connection \eqref{fteqspin} coincide with the antisymmetric part of the field equations for the tetrad \eqref{fteqas}.

We notice that a very special situation occurs when we have $\Tw=T_0=\text{const}$. In this case the spin connection field equation \eqref{fteqspin} is automatically satisfied and the tetrad field equations \eqref{ftequationG} reduce to the ordinary Einstein equations with the effective cosmological constant $(f(T_0)-T_0f_T(T_0))/2$. This means that if we are able to construct a constant torsion scalar $\Tw=T_0$ for some solution in general relativity, then this solution remains a solution of $f(T)$ gravity for an arbitrary function $f$.  Using this method it was shown that the well known Schwarzschild~\cite{Ferraro:2011ks}, Kerr~\cite{Bejarano:2014bca} and McVittie~\cite{Bejarano:2017akj} solutions of general relativity solve $f(T)$ gravity as well.

\subsection{Local Lorentz symmetry in \texorpdfstring{$f(T)$}{f(T)} gravity}

The issue that has caused a lot of attention and raised some doubts about the consistency of $f(T)$ gravity and other modified teleparallel gravity models is the question of local Lorentz invariance~\cite{Sotiriou:2008rp,Sotiriou_Li_Barrow2011}. We will now explain the origins of this  problem and how it is avoided in the covariant formulation of the theory that we use here.

In Section~\ref{secPT} we have mentioned the so-called pure tetrad approach to teleparallel gravity, where the only variable is the tetrad. This originates from the fact that the teleparallel connection \eqref{InerConn} is a pure gauge connection and hence it is always possible to perform a local Lorentz transformation such that the connection is transformed to zero. This is then equivalent to choosing a specific frame in which the spin connection vanishes and hence formulating the theory in this very specific class of frames. Strictly speaking, the question of local Lorentz invariance is ill-defined in this case since we choose the specific frame and hence we are not allowed to perform local Lorentz transformations.

Nevertheless, in the case of the ordinary teleparallel gravity with Lagrangian density \eqref{TeleLagra} this approach gained some popularity not only because it was originally used in Einstein's teleparallelism~\cite{Einstein1928,Sauer:2004hj} but also since it can be justified in some cases. Particularly, on account of property \eqref{rel}, the spin connection does not dynamically affect the field equations \eqref{tfe10} and hence setting it to zero can be used to obtain solutions. In fact, as we have explained in Section~\ref{solutions}, in our covariant approach to ordinary teleparallel gravity, for the sake of convenience we can also set the spin connection to zero as an intermediate step in our calculations.   The issue of local Lorentz invariance manifests itself only when problems beyond the solutions are considered; i.e., the total value of the action and its renormalization, definition of the energy-momentum, etc.

The problem of the original formulation of $f(T)$ gravity with Lagrangian density \eqref{ftaction} is due to the fact that the theory was originally formulated as a modification of the pure tetrad teleparallel gravity.  The problem of local Lorentz symmetry violation was then inherited by the $f(T)$ gravity in a more serious way since it also affects the solutions of the field equations. In particular, this  leads to the situation where only some frames in the same equivalence class (i.e., corresponding to the same metric), were able to solve the field equations. Such frames were referred to as \emph{good tetrads}~\cite{Tamanini_Boehmer2012}, while the so-called \emph{bad tetrads}--related to good tetrads by a local Lorentz transformation--were solutions only in the limit of the ordinary teleparallel gravity. For example, in the case of spherical symmetry, the diagonal tetrad \eqref{tetdiag} was considered to be a \emph{bad tetrad}, while the off-diagonal \eqref{tetprop} was a \emph{good tetrad}.

It is easy to understand this problem of the original $f(T)$ gravity from the viewpoint of our covariant formulation. As we have argued in the previous Section~\ref{secfteq}, the field equations of $f(T)$ gravity determine both the tetrad and the spin connection, and hence the solution is always a pair of variables $\{h^a{}_{\mu},\omegaw^a{}_{b\mu}\}$. Since these variables are not independent, a transformation of the spin connection must be always accompanied by a transformation of the corresponding tetrad.
In particular, if we transform the spin connection to zero, then we must perform a simultaneous transformation on the tetrad; i.e.,
\begin{equation}
\{h^a_{\ \mu},\omegaw^a_{\ b\mu}\} \xrightarrow{\hspace*{0.7cm}} \{\tilde{h}^a_{\ \mu},0\}. \label{wtransf2}
\end{equation}
We can then identify the class of frames $\tilde{h}^a_{\ \mu}$ which corresponds to the zero connection within the class of good frames. However, we can see that there is nothing fundamentally special about these frames. All other frames related by local Lorentz transformation are equally viable; we just need to use  the corresponding spin connections with them.

At least in principle, it is possible to write down the field equations directly in terms of the good frames $\tilde{h}^a_{\ \mu}$, which are obtained by  using the transformation \eqref{wtransf2} on the field equations \eqref{ftequation}. As a result, we will obtain 16 field equations that completely determine the tetrad, including the local Lorentz degrees freedom of the tetrad. This is in contrast with the situation in the ordinary teleparallel gravity discussed in Section~\ref{solutions} where the field equations determined only the equivalence class of the tetrads. The tetrad that solves the $f(T)$ field equations after the transformation \eqref{wtransf2} is completely fixed.

However, this does not imply that local Lorentz symmetry is violated in $f(T)$ gravity because we need to keep in mind that we have obtained this solution using the transformation \eqref{wtransf2} and hence we work only in this class of frames. If we want to discuss local Lorentz symmetry we need to act with local Lorentz transformations on both variables which necessarily generates a new spin connection. We can picture this as having  $\tilde{h}^a_{\ \mu}$ that solves the field equations with a zero spin connection, and then we perform an inverse of a local Lorentz transformation \eqref{wtransf2} with an arbitrary $\Lambda^a_{\ b}$. This will then generate for each tetrad $h^a_{\ b}$ a corresponding spin connection $\omegaw^a_{\ b\mu}$.

\subsection{Examples: Solutions in \texorpdfstring{$f(T)$}{f(T)} gravity}

\subsubsection{Example: Minkowski spacetime\label{secftex1}}

A rather trivial but very illustrative example of the relevance of the spin connection and the problems of the original formulation of $f(T)$ gravity is the simple Minkowski spacetime. We consider two different tetrads representing the Minkowski spacetime. The first one is the diagonal tetrad in the Cartesian coordinate system:
\begin{equation}\label{tetminkcart}
h^a_{\ \mu}=\text{diag}\left(1,1,1,1\right),
\end{equation}
which--if we set the spin connection to zero--leads to zero torsion and hence for any function $f(T)$ with $f(0)=0$ then the field equations are trivially satisfied. Therefore, the tetrad \eqref{tetminkcart} is a proper tetrad to which corresponds a vanishing zero connection or, in the terminology of~\cite{Tamanini_Boehmer2012}, a good tetrad.

On the other hand, if we consider a Minkowski diagonal tetrad in the spherical coordinate system
\begin{equation}\label{tetminkrad}
h^a_{\ \mu}=\text{diag}\left(1,1,r,r\sin\theta\right),
\end{equation}
and use it with the vanishing zero spin connection, we find that the corresponding torsion scalar is non-zero. Additionally, we observe that one of the field equations,
\begin{equation}
E_{\hat{2}}^{\ \theta}=\frac{4 h f_{TT}\cot\theta}{r^{5}}=0,
\end{equation}
has a solution only if we set $f_{TT}=0$. This reduces the theory back to the ordinary teleparallel gravity and hence \eqref{tetminkrad} is, in the terminology of~\cite{Tamanini_Boehmer2012}, a bad tetrad.

The problem of the original formulation of $f(T)$ gravity was that in order to be able to use the spherical coordinate system in $f(T)$ gravity to describe Minkowski spacetime we had to use the tetrad $\tilde{h}^a_{\ \mu}=\tilde{\Lambda}^a_{\ b} h^a_{\ \mu}$ with $\tilde{\Lambda}^a_{\ b}$ given by \eqref{specLorentz} and $h^a_{\ \mu}$ given by \eqref{tetminkrad}, which yields the equivalent of the reference tetrad in Eq.~\eqref{tetpropref}.  Since gravity is absent in  Minkowski spacetime, this demonstrates very well that the nature of the Lorentz invariance problem is not related to the modification of gravity but is just a matter of the consistent formulation of the theory.
On the other hand, in the covariant formulation of $f(T)$ gravity, the problem is absent and we are able to use all tetrads in arbitrary coordinate systems. In the case of Minkowski spacetime, for instance, we are free to use the diagonal tetrad \eqref{tetminkrad} but we need to use it along with its corresponding spin connection \eqref{spschw}.

\subsubsection{Example: FLRW spacetime\label{secftex2}}

We now move on to the more non-trivial example of the  FLRW spacetime with zero spatial curvature describing the evolution of the Universe. Using the Cartesian coordinate system we can choose a diagonal tetrad  as,
\begin{equation}
h^a_{\ \mu}=\text{diag} (1,a(t),a(t),a(t)), \label{tetradfrwc}
\end{equation}
with $a(t)$  being the scale factor. This tetrad leads to the torsion scalar
\begin{equation}
\Tw=-6H^2,
\end{equation}
where $H=\dot{a}/a$ is the Hubble parameter, and gives rise to the Friedmann equations
\begin{align}
\kappa \rho_M &= 6 H^2 f_T +\frac{1}{2}f, \label{fried1}\\
\kappa(p_M+\rho_M) &=2 \dot{H}(12 f_{TT}H^2-f_T),\label{fried2}
\end{align}
where $\rho_M$ and $p_M$ are the energy density and pressure of the matter fluid, respectively. These are the correct $f(T)$ modified Friedmann equations capable of explaining the accelerated expansion of the Universe, as shown originally in~\cite{Ferraro:2006jd,Ferraro:2008ey,Bengochea:2008gz,Linder:2010py} and later extensively studied in~\cite{Chen:2010va,Wu:2010mn,Zheng:2010am,Bamba:2010wb,Wu:2011kh,Dent:2011zz,Wei:2011aa,Atazadeh:2011aa,Cai:2011tc,Capozziello:2011hj,Farajollahi:2011af,Cardone:2012xq,Izumi:2012qj,Amoros:2013nxa,Nesseris:2013jea,Harko:2014sja,Haro:2014wha,Capozziello:2015rda,Nashed:2015pda,Oikonomou:2016jjh,Oikonomou:2016jjh,Wu:2016dkt,Farrugia:2016qqe,Nunes:2016qyp,Nunes:2016plz,Skugoreva:2017vde}. For a complete list of references see the review~\cite{Capozziello:2017bxm}.

The observation that the tetrad \eqref{tetradfrwc} leads directly to some interesting dynamics means that this tetrad is already in  the proper form and hence leads to symmetric field equations using the  zero connection. Let us mention that at the perturbative level one has to consider perturbations to all components of the tetrad \eqref{tetradfrwc} and solve the antisymmetric part of the field equations, which are  non-trivial even for the Cartesian tetrad \eqref{tetradfrwc} at the perturbative level, in order to obtain the correct cosmological perturbation theory~\cite{Golovnev:2018wbh}.

Similar to the case of Minkowski spacetime, if we would like to use instead of the Cartesian diagonal tetrad \eqref{tetradfrwc} the diagonal tetrad in the spherical coordinate system
\begin{equation}
h^a_{\ \mu}=\text{diag} (1, a(t), a(t)\,r, a(t)\,r\sin\theta), \label{tetradfrwr}
\end{equation}
we have to either use the spin connection corresponding to \eqref{tetradfrwr}, which is given by \eqref{spschw}, or make the tetrad \eqref{tetradfrwr} proper by the local Lorentz transformation given by \eqref{specLorentz}. Both methods will lead to the same set of Friedmann equations \eqref{fried1}-\eqref{fried2} as when working with the Cartesian tetrad \eqref{tetradfrwc}.

It is also interesting to consider spatially non-flat FLRW spacetimes represented by the tetrad in the spherical coordinate system
\begin{equation}
h^a_{\ \mu}=\text{diag} \left(1, \frac{a(t)}{\sqrt{\chi}}, a(t)\,r, a(t)\,r\sin\theta\right), \label{tetradflrwcurv},
\end{equation}
where $\chi=1-kr^2$.

For the positively spatially curved FLRW spacetime, $k=+1$, the teleparallel spin connnection is given by ~\cite{Hohmann:2019nat}
\[\begin{array}{llll}%
& \quad \omw_{\pm}^{\hat{1}}{}_{\hat{2}\theta} = -\omw_{\pm}^{\hat{2}}{}_{\hat{1}\theta} = -\chi\,,
& \quad \omw_{\pm}^{\hat{1}}{}_{\hat{2}\phi} = -\omw_{\pm}^{\hat{2}}{}_{\hat{1}\phi} = \pm r\sin\theta\,, \quad
& \quad \omw_{\pm}^{\hat{1}}{}_{\hat{3}\theta} = -\omw_{\pm}^{\hat{3}}{}_{\hat{1}\theta} = \mp r\,, \quad \nonumber 
\\
& \quad\omw_{\pm}^{\hat{1}}{}_{\hat{3}\phi} = -\omw_{\pm}^{\hat{3}}{}_{\hat{1}\phi} = -\chi\sin\theta\,, 
& \quad\omw_{\pm}^{\hat{2}}{}_{\hat{3}r} = -\omw_{\pm}^{\hat{3}}{}_{\hat{2}r} = \pm\frac{1}{\chi}\,, 
& \quad \omw_{\pm}^{\hat{2}}{}_{\hat{3}\phi} = -\omw_{\pm}^{\hat{3}}{}_{\hat{2}\phi} = -\cos\theta\,,\label{spinpos}\end{array}
\]
where $\pm$ represents two possible sign choices.

In the case of  the negatively spatially curved FLRW spacetime, the teleparallel spin connection is given by~\cite{Hohmann:2019nat}
\[\begin{array}{llll}%
& \quad \omw^{\hat{0}}{}_{\hat{1}r} = \omw^{\hat{1}}{}_{\hat{0}r} = \frac{1}{\chi}\,, 
& \quad \omw^{\hat{0}}{}_{\hat{2}\theta} = \omw^{\hat{2}}{}_{\hat{0}\theta} = r\,, 
& \quad \omw^{\hat{0}}{}_{\hat{3}\phi} = \omw^{\hat{3}}{}_{\hat{0}\phi} = r\sin\theta\,, \nonumber\\
& \quad \omw^{\hat{1}}{}_{\hat{2}\theta} = -\omw^{\hat{2}}{}_{\hat{1}\theta} = -\chi\,, 
& \quad \omw^{\hat{1}}{}_{\hat{3}\phi} = -\omw^{\hat{3}}{}_{\hat{1}\phi} = -\chi\sin\theta\,, 
& \quad \omw^{\hat{2}}{}_{\hat{3}\phi} = -\omw^{\hat{3}}{}_{\hat{2}\phi} = -\cos\theta\,. \label{spinneg}
\end{array}
\]
Both results \eqref{spinpos} and \eqref{spinneg} can be transformed using the corresponding local Lorentz tranformations to the proper tetrad-form where the spin connection is zero and the tetrad \eqref{tetradflrwcurv}  takes a non-diagonal form, see~\cite{Hohmann:2019nat}. Moreover, in the case the negatively spatially curved FLRW spacetime, there seems to exists also a complex spin connection~\cite{Hohmann:2019nat} corresponding to the complex tetrad first examined in~\cite{Capozziello:2018hly}.

\subsubsection{Example: Spherically symmetric vacuum spacetime\label{secftex3}}

The spherically symmetric solutions of the field equations in any gravitational theory are of the crucial importance since they describe the gravitational field outside a massive spherical body which is important for understanding the 	dynamics of the solar system and the dynamics of more exotic objects such as black holes.  In the framework of $f(T)$ gravity, this problem was considered in
\cite{Ferraro:2011us,Ferraro:2011ks,Boehmer:2011gw,Bohmer:2011si,Tamanini_Boehmer2012,Wang:2011xf,HamaniDaouda:2011iy,Dong:2012en,delaCruz-Dombriz:2014zaa,Junior:2015dga,Das:2015gwa,Zubair:2015cpa,Ruggiero:2015oka}.

The spherically symmetric metric (see also Eq.~\eqref{schwmet}) has the form
\begin{equation}
ds^2=A(r)^2 dt^2-B(r)^2 dr^2-r^2d\theta^2-r^2\sin^2\theta d\phi^2\,.
\label{metricspherical}
\end{equation}
The most natural choice of the tetrad corresponding to this metric has the simple diagonal form (see also in Eq.~\eqref{tetdiag}), the same choice as in the ordinary teleparallel gravity,
\begin{equation}
h^a_{\ \mu}=\text{diag}\left(A,B,r,r\sin\theta\right)
\label{schwtet1}.
\end{equation}	
Same as in the case of Minkowski vacuum in the spherical coordinate system \eqref{tetminkrad}, this diagonal tetrad leads to the field equations that can be satisfied only in the case when $f(T)$ gravity reduces to the ordinary teleparallel gravity. To see this, we can straightforwardly check that the diagonal tetrad \eqref{tetdiag} with a trivial spin connection $\omega^a_{\ b\mu}=0$, gives us the torsion scalar
\begin{equation}
\Tw=\frac{2(A+rA')}{r^2 A B^2},
\end{equation}
where the prime denotes  the derivative of the torsion scalar \eqref{Tor-Spherical-Sym} with respect to the coordinate $r$. One of the field equations is then given by
\begin{equation}\label{probeq}
E_{\hat{2}}^{\ \theta}=\frac{hf_{TT}\cot\theta}{Br}  \Tw'=0,
\end{equation}
which can be satisfied only if $f_{TT}=0$, and hence restricts the theory to the ordinary teleparallel gravity. Due to this fact, the diagonal tetrad \eqref{schwtet1} was considered to be not a consistent solution in  the original (non-covariant) formulation of $f(T)$ gravity.

We now demonstrate that this issue is not present in the covariant formulation  and that it is possible to obtain non-trivial solutions using an arbitrary tetrad corresponding to the metric \eqref{metricspherical} provided that it is accompanied by the corresponding spin connection. The most straighforward method to determine the spin connection is to solve its field equations \eqref{fteqspin}, but this it is a  difficult task to achieve in practice. Therefore,  we can use the method of reference tetrads described in Section~\ref{seccon}. The only difference compared to the ordinary teleparallel gravity is that we do not know the solution for the tetrad and hence we cannot  straightforwardly determine the reference tetrad using  \eqref{reftet}. Nevertheless, it is reasonable to  assume that in the absence of gravity, the diagonal tetrad \eqref{schwtet1} should reduce to the tetrad \eqref{tetminkrad} representing the Minkowski spacetime in spherical coordinates and hence the corresponding spin connection is given by (see also Eq.~\eqref{spschw})
\begin{equation}
\omw^{\hat{1}}_{\ \hat{2}\theta}=-\omw^{\hat{2}}_{\ \hat{1}\theta}=-1, \quad
\omw^{\hat{1}}_{\ \hat{3}\phi}=-\omw^{\hat{3}}_{\ \hat{1}\phi}=-\sin\theta, \quad
\omw^{\hat{2}}_{\ \hat{3}\phi}=-\omw^{\hat{3}}_{\ \hat{2}\phi}=-\cos\theta. \label{spschw1}
\end{equation}
We can check that the field equations for the spin connection \eqref{fteqspin} are indeed satisfied for the tetrad \eqref{schwtet1} and the spin connection \eqref{spschw1}.
We find then that the torsion scalar
\begin{equation}
\Tw(h^a_{\ \mu},\omega^a_{\ b\mu})=-\frac{2 (B-1) \left(A-A B+2 r A'\right)}{r^2 A B^2},\label{Tor-Spherical-Sym}
\end{equation}
and the field equations \eqref{ftequation}  are given by
\begin{align}
E_{\hat{0}}^{\ t}&\equiv h\Bigg(
\frac{1}{2A}f
+
2{f_T} \frac {\left(-AB+AB^2+AB'r+A'B^2r- A'Br \right) }{ A^2B^3r^2 }+ 2{f_{TT}}\Tw' \frac {(B -1)}{ AB^2r} \Bigg)\,,
\label{eq00}\\
E_{\hat{1}}^{\ r}&\equiv h\Bigg( \frac{1}{2B}f
+
2{f_T} \frac {\left(-2A'r+AB+A'Br-A \right) }{ AB^3r^2 }\Bigg)\,,
\\
E_{\hat{2}}^{\ \theta}&\equiv h\Bigg( \frac{1}{2r}f
+
{f_T} \frac {\left(2B^2A'r-AB-AB^3+2AB^2-A''Br^2+AB'r-3A'Br+B'A'r^2 \right) }{ AB^3r^3 }
\nonumber\\
&\qquad\qquad + {f_{TT}}\Tw' \frac {(AB -A-A'r)}{ AB^2r^2} \Bigg)\,,
\\
E_{\hat{3}}^{\ \phi}&\equiv \frac{1}{\sin\theta} E_{\hat{2}}^{\ \theta}\,.
\label{eq33}
\end{align}

We can observe that the field equations \eqref{eq00}-\eqref{eq33} do not restrict the form of the $f(T)$-function, in contrast with \eqref{probeq}, and hence generally lead to  new solutions distinct from the ordinary teleparallel gravity. Moreover, we can now  use  any tetrad corresponding to the metric \eqref{metricspherical} provided that the corresponding spin connection is calculated. We can check that the off-diagonal tetrad \eqref{tetprop} with the zero spin connection lead to the same field equations \eqref{eq00}-\eqref{eq33}. However, the tetrad \eqref{tetprop} does not have any priviliged position in the covariant formulation of $f(T)$ gravity; it is just a specific tetrad in which the corresponding spin connection happens to be zero.


\section{Other Modified Teleparallel Models}\label{sec:Other-Modified}

The most popular modified teleparallel gravity model in the recent decade is $f(T)$ gravity and much of the attention has been focused on this particular model. However, the teleparallel structure allows us to formulate a plethora of other interesting modified teleparallel gravity models. We briefly review here some of the more popular models and provide a classification scheme based on their essential features and/or the purpose for which they were proposed.

\subsection{New General Relativity\label{sec:NGR}}

The so-called New General Relativity, introduced in 1979 by Hayashi and Shirafuji \cite{Hayashi_Shirafuji1979,Hayashi_Shirafuji1982},  is the oldest modified teleparallel gravity model where the teleparallel Lagrangian \eqref{TeleLagra} is straightforwardly generalized by considering arbitrary coefficients of the quadratic scalar torsion terms.
We follow here the  original approach used in \cite{Hayashi_Shirafuji1979,Hayashi_Shirafuji1982}, which will be also useful later in Section~\ref{secFTTT}, and decompose the torsion tensor into irreducible parts with respect to the Lorentz group
\begin{align}
\Tw_{\lambda\mu\nu} = \frac{2}{3}(t_{\lambda\mu\nu}-t_{\lambda\nu\mu}) +
\frac{1}{3}(g_{\lambda\mu}v_{\nu}-g_{\lambda\nu}v_{\mu}) +
\epsilon_{\lambda\mu\nu\rho}a^{\rho}\,,
\end{align}
where
\begin{align}
v_{\mu} &= \Tw^{\lambda}{}_{\lambda\mu}\,,\\
a_{\mu} &= \frac{1}{6}\epsilon_{\mu\nu\sigma\rho}\Tw^{\nu\sigma\rho}\,,\\
t_{\lambda\mu\nu} &= \frac{1}{2}(\Tw_{\lambda\mu\nu}+\Tw_{\mu\lambda\nu}) +
\frac{1}{6}(g_{\nu\lambda}v_{\mu}+g_{\nu\mu}v_{\lambda})-\frac{1}{3}g_{\lambda\mu}v_{\nu}\,,
\end{align}
are known as the vector, axial, and purely tensorial torsions, respectively. We can then construct three parity preserving quadratic invariants
\begin{align}
T_{\rm ten} &= t_{\lambda\mu\nu}t^{\lambda\mu\nu} =
\frac{1}{2}\Big(\Tw_{\lambda\mu\nu}\Tw^{\lambda\mu\nu}+\Tw_{\lambda\mu\nu}\Tw^{\mu\lambda\nu}\Big)-\frac{1}{2}\Tw^\lambda{}_{\lambda\mu}\Tw_\nu{}^{\nu\mu}\,,
\label{Tten}\\
T_{\rm ax} &= a_{\mu}a^{\mu} =
\frac{1}{18}\Big(\Tw_{\lambda\mu\nu}\Tw^{\lambda\mu\nu}-2\Tw_{\lambda\mu\nu}\Tw^{\mu\lambda\nu}\Big)\,,
\label{Tax}\\
T_{\rm vec} &= v_{\mu}v^{\mu} =
\Tw^\lambda{}_{\lambda\mu}\Tw_\nu{}^{\nu\mu}
\label{Tvec}\,.
\end{align}
The action of the teleparallel gravity \eqref{TeleLagra} in terms of these quadratic invariants takes the form
\begin{equation}
\Lw_\text{TG} =
{\frac{h}{16 \pi G}} \left(\frac{3}{2} T_{\rm ax} + \frac{2}{3} T_{\rm ten}  
- \frac{2}{3} T_{\rm vec}\right).
\label{NGRLag}
\end{equation}

New General Relativity is a straightforward generalization of the teleparallel Lagrangian where the coefficients in front of the quadratic invariants take arbitrary values; i.e.,
\begin{equation}
\Lw_{\rm NGR} =
\frac{h}{2\kappa}\Big(a_{0} + a_{1} T_{\rm ax} + a_{2} T_{\rm ten} + a_{3} T_{\rm vec} \Big) \,,
\label{action1}
\end{equation}
where the four $a_{i}$ are arbitrary constants and $a_{0}$ can be interpreted as the cosmological constant.  Since, up to a divergence,
\begin{equation}
- \frac{3}{2} T_{\rm ax} - \frac{2}{3} T_{\rm ten} + \frac{2}{3} T_{\rm vec} = - \Rbol \,,
\end{equation}
with $\Rbol$ the scalar curvature of the Levi-Civita connection, Hayashi and Shirafuji rewrote the Lagrangian \eqref{action1} in the form
\begin{equation}
\Lw_{\rm NGR} =
\frac{h}{2\kappa}\Big(a_{0} - \Rbol + b_{1} T_{\rm ax} + b_{2} T_{\rm ten} + b_{3} T_{\rm vec} \Big) \,,
\label{action1bis}
\end{equation}
with the new coefficients given by
\begin{equation}
b_1 = a_1 + \frac{2}{3}, \quad b_2 = a_2 + \frac{2}{3}, \quad b_3 = b_3 - \frac{3}{2} \,.
\end{equation}
In this theory,  torsion would represent additional degrees of freedom relative to the curvature, which would thus produce deviations in relation to general relativity, or equivalently, in relation to teleparallel gravity. In the original new general relativity by Hayashi and Shirafuji \cite{Hayashi_Shirafuji1979,Hayashi_Shirafuji1982}, only the $b_1$ parameter was considered to be  non zero since solar system experiments put strong constraints on $b_2$ and $b_3$. Further problems and limitations of this model were discussed in \cite{Kopczynski1982,Nester1991,Cheng:1988zg}.

\subsection{Conformal Teleparallel Gravity}
Recently there has been increased interest in gravitational theories with conformal invariance, which is expected to be recovered as a fundamental symmetry at the Planck scale  \cite{Mannheim:2011ds,Maldacena:2011mk,Hooft:2014daa,Anastasiou:2016jix}.  In the standard Riemannian framework the Lagrangian is usually assumed to be quadratic in the Weyl tensor and hence leads to a theory with fourth order field equations.

Within the teleparallel framework we can use the fact that the torsion contains only first derivatives of the tetrad and construct a conformally invariant theory of gravity with second order field equations \cite{Maluf:2011kf}. The most general Lagrangian of the \emph{teleparallel conformal gravity} is then given by
\begin{equation}\label{LagCTG}
\Lw_{TCG}= h L_1 L_2,
\end{equation}
where $L_1$ is a generalization of the torsion scalar \eqref{Tscalar} given by
\begin{equation}
L_1=a_1\, \Tw^\rho{}_{\mu\nu} \, \Tw_\rho{}^{\mu \nu} + a_2\, \Tw^\rho{}_{\mu\nu} \, \Tw^{\nu \mu}{}_\rho + a_3\, \Tw^\rho{}_{\mu\rho} \, \Tw^{\nu \mu}{}_\nu.
\label{TscalarCTG}
\end{equation}
where $a_1,a_2,a_3$ are three constants satisfying the relation
\begin{equation}\label{CTGcond}
2a_1+a_2+3a_3=0,
\end{equation}
and $L_2$ is defined analogously to \eqref{TscalarCTG} with generally three different constants $a'_1,a'_2,a'_3$ satisfying the analogous constraint to \eqref{CTGcond}.

Using the quadratic invariants of the irreducible parts of the torsion tensor \eqref{Tten}-\eqref{Tvec}, it is possible to write the torsion scalar \eqref{TscalarCTG} in a simpler form 
\begin{equation}
L_1=b_1 T_{ax} + b_2 T_{ten},
\label{TscalarCTG1}
\end{equation}
where $b_1, b_2$ are arbitrary constants \cite{Bahamonde:2017wwk}. The second scalar $L_2$ can be written analogously with different constants $b'_1, b'_2$. Since the overall normalization fixes one of the constants, the teleparallel conformal model \eqref{LagCTG} has 3 free parameters that can be chosen arbitrarily. 

See \cite{Iosifidis:2018zwo} for a recent generalization of conformal teleparallel gravity. Alternatively,  teleparallel conformal gravity can be realized by coupling a conformal scalar field to the generalized torsion scalar \eqref{TscalarCTG} \cite{Maluf:2011kf,Bamba:2013jqa,daSilva:2017xks}, or using Kaluza-Klein reduction \cite{Geng:2014nfa,Geng:2016yke}.

\subsection{\texorpdfstring{$f(T_\text{ax},T_\text{ten},T_\text{vec})$}{f(Tax,Tten,Tvec)} Gravity \label{secFTTT}}

A natural generalization combining elements of both $f(T)$ gravity and new general relativity is $f(T_\text{ax},T_\text{ten},T_\text{vec})$ gravity \cite{Bahamonde:2017wwk},  where the Lagrangian is taken to be an arbitrary function of quadratic invariants of the irreducible parts of the torsion tensor \eqref{Tten}-\eqref{Tvec}
\begin{align}
\Lw_{ATV} = \frac{h}{2\kappa} \,
f(T_{\rm ax},T_{\rm ten},T_{\rm vec}).
\label{action3}
\end{align}
This model includes other models discussed in previous sections as special cases and is particularly suitable to study the behavior of teleparallel models  under conformal transformations of the metric $\hat{g}_{\mu\nu}=\Omega^2(x) g_{\mu\nu}$, or
equally of the tetrad $\hat{h}^a_{\phantom{a}\mu} =\Omega(x) h^a_{\phantom{a}\mu} $, where $\Omega(x)$ is the conformal factor. The quadratic invariants of the irreducible parts of the torsion tensor \eqref{Tten}-\eqref{Tvec}  transform as
\begin{align}\label{cttor1}
T_{\rm ax} = \Omega^2 \hat{T}_{\rm ax}, \qquad
T_{\rm ten} = \Omega^2 \hat{T}_{\rm ten}, \qquad
T_{\rm vec} = \Omega^2\hat{T}_{\rm vec}+6\Omega\hat{v}^{\mu}\hat{\partial}_{\mu}\Omega+9\hat{g}^{\mu\nu}(\hat{\partial}_{\mu}\Omega)(\hat{\partial}_{\nu}\Omega).
\end{align}
Analyzing these transformation properties it can then be concluded,
unlike the situation in $f(R)$ gravity \cite{Sotiriou:2008rp},
that it is not possible  to find an ``Einstein frame" where the theory reduces to ordinary general relativity and a minimally coupled scalar field \cite{Bahamonde:2017wwk}. This is a generalization of the result previously obtained in $f(T)$ gravity \cite{Yang:2010ji,Wright:2016ayu}. Moreover, we can also quickly confirm the observation that \eqref{TscalarCTG1} transforms properly under conformal transformations and hence the Lagrangian \eqref{LagCTG} is, indeed, conformally invariant.

\subsection{Gravity Models Inspired by Axiomatic Electrodynamics}

A novel approach to teleparallel theories was recently proposed by Itin \emph{et al.} \cite{Itin:2016nxk}  that utilizes the similarities between electromagnetism and teleparallel gravity. The field equations of electromagnetism can be written as
\begin{align}
\mathrm{d} F &= 0,  \label{eq1} \\ \mathrm{d} H &= J, \label{eq2}
\end{align}
where $F$ is the electromagnetic field strength $2$-form, $H$ is the excitation $2$-form and $J$ the current $3$-form. The specific case of Maxwell electrodynamics is then defined by the relation between  the excitation form and the field strength form using the Hodge dual map
\begin{align}\label{MaxED}
H=  \star F\,.
\end{align}
The crucial observation  of the so-called \emph{axiomatic electrodynamics}   is that the field equations \eqref{eq1}-\eqref{eq2} describe a consistent theory of electrodynamics even if the  relation \eqref{MaxED} is generalized to a more general \emph{constitutive relation} $H=\kappa(F)$. Using this generalization we can then describe various effects in media and  non-linear theories of electrodynamics in one unified language \cite{Hehl}.  

Teleparallel gravity can be cast into a similar form by writing the Lagrangian in the language of differential forms as
\begin{align}\label{eq:TEGRAction}
\L = \Tw^a \wedge \Hw_a,
\end{align}
where $\Hw_a$ is the gravitational excitation 2-form related to the  superpotential \eqref{sup} through
\begin{equation}\label{CRTEGR}
\Hw_a(h^b, T^b) =\frac{1}{4} h \epsilon_{\mu\nu\rho\sigma} \sw_a{}^{\rho\sigma} d x^\mu \wedge  d x^\nu\,.
\end{equation}
The Bianchi identities and the field equations of teleparallel gravity then take a form that closely resembles the equations of electrodynamics \eqref{eq1}-\eqref{eq2}:
\begin{align}\label{FirstEq}
\D \Tw^a &= 0\,,\\
\D\Piw_{a} - \Upsilonw_{a}& = \Sigmaw_{a}\,,\label{SecondEq}
\end{align}
where $\D$ is the teleparallel covariant exterior derivative, and $\Upsilonw_{a}$ and $\Sigmaw_{a}$ are the  gravitational and matter energy-momentum 3-forms defined in  \cite{Hohmann:2017duq}.  In the case of ordinary teleparallel gravity we have that $\Piw_{a}=\Hw_a$, and using the so-called generalized Hodge dual \cite{Lucas:2008gs} the excitation form $\Hw_a$ can be related to a torsion form in a similar way as in the Maxwell case \eqref{MaxED}.

In \cite{Itin:2016nxk} it was demonstrated that the same generalization to that in the case of axiomatic electrodynamics can be realized  in teleparallel gravity by replacing the constitutive relation \eqref{CRTEGR} by a general constitutive relation
\begin{equation}
\label{CRgen}
\Hw_a=\Hw_a(h^b,\Tw^b),
\end{equation}
where $\Hw_a(h^b,\Tw^b)$ is a function of the tetrad and the torsion. It was shown that the most general local and linear  constitutive relation defines new general relativity discussed in Section~\ref{sec:NGR} \cite{Itin:2016nxk,Itin:2018lcb,Itin:2018dru}. In \cite{Hohmann:2017duq} this approach  was generalized to include arbitrary non-linear constitutive relations and it was shown that all previously studied modified teleparallel gravity models with second order field equations can be naturally realized by finding corresponding constitutive relations. For example, $f(T)$ gravity can be realized through the constitutive relation
\begin{equation}
\Hw_{a}(h^a, T^a)= \frac{1}{4} h \frac{f(\Tw)}{\Tw}  \epsilon_{\mu\nu\rho\sigma} \sw_a{}^{\rho\sigma}d x^\mu \wedge  d x^\nu\,.
\end{equation}
The advantage of such an axiomatic approach is that the field equations of all modified teleparallel theories then take the same form \eqref{FirstEq}-\eqref{SecondEq}, which allows us to analyze them in full generality and  better understand their generic features. For instance, in \cite{Hohmann:2017duq} it was shown that the result from $f(T)$ gravity, discussed in Section~\ref{secfteq}, that the field equations for the spin connection coincide with the antisymmetric part of the field equations for the tetrad, applies to all modified gravity models with second order field equations.  Moreover,  we obtain also a new framework where  modified gravity theories are viewed analogously  to various non-linear electrodynamics theories. Using such analogies we can then construct new modified gravity models inspired by electrodynamics theories \cite{Hohmann:2017duq}.

\subsection{Teleparallel Dark Energy and Scalar-Torsion Models}

Another popular way to modify gravity is by introducing   scalar fields. In the standard curvature-based approach, the earliest such model was Brans-Dicke gravity \cite{Brans:1961sx}, and more recent models include various quintessence and scalar-tensor gravity models \cite{Ratra:1987rm,Wetterich:1987fm,Caldwell:1997ii,Fujii:2003pa}.

In the teleparallel framework, we can follow the same path and formulate modified gravity theories with scalar fields.  The first model introduced by Geng \emph{et al.} \cite{Geng:2011aj} under the name  \emph{teleparallel dark energy} considers the torsion scalar \eqref{Tscalar} non-minimally coupled to the scalar field with kinetic and potential terms for the scalar field:
\begin{equation}
\Lw_{\rm TDE} =
h \Big[ \frac{\Tw}{2\kappa}  +\frac{1}{2}\xi \Tw \phi^2 + \frac{1}{2} (\partial_\mu \phi) (\partial^\mu \phi) -V(\phi) \Big] \,.
\label{actionTDE}
\end{equation}
It was shown that such a model leads to interesting cosmological dynamics consistent with observations yet distinct from $f(T)$ gravity or curvature-based analogues \cite{Geng:2011ka,Wei:2011yr,Wu:2011kh,Xu:2012jf,Gu:2012ww,Sadjadi:2013nb,Skugoreva:2014ena,Jarv:2015odu}. Note that this model, in the same way as $f(T)$ gravity discussed in Section~\ref{secfT}, was originally presented as violating local Lorentz invariance, but it  can be reformulated in a covariant way using the non-trivial spin connection \cite{Hohmann:2018rwf}.

Following this example other modified gravity models with a scalar field were proposed, including the possibility of coupling the gradient of the scalar field with the trace of the torsion tensor \cite{Otalora:2014aoa}, and a tachyonic scalar field \cite{Banijamali:2012nx}. Recently, the most general extension was proposed as $f(T,X,Y,\phi)$, where $X$ is the kinetic term for the scalar field, and $Y$ is the term representing the coupling between the torsion and gradient of the scalar field \cite{Hohmann:2018vle,Hohmann:2018dqh,Hohmann:2018ijr}. A similar yet distinct  model is  $f(T,\mathcal{T})$  gravity \cite{Harko:2014aja}, where the scalar $\mathcal{T}=\Theta^\mu{}_{\mu}$  is the trace of the energy-momentum of matter \eqref{matemten}.

\subsection{Higher Derivative Models: \texorpdfstring{$f(T,B)$}{f(T,B)} and Others\label{secModHD}}

All of the previous teleparallel models considered the Lagrangian to be a function of the torsion scalar only and did not include its derivatives; as a result the equations of motion were always second order. However, it is possible to include derivatives of torsion and deal with fourth (or possibly higher) order equations. Among such higher derivative type models, the best motivated one is perhaps $f(T,B)$ gravity,
\begin{equation}
\Lw_{\rm fTB} =
\frac{h}{2\kappa} f(\Tw,\Bw)
\label{actionfTB}
\end{equation}
where we include the `boundary' term $\Bw=(2/e)\, \partial_\mu (e v^\mu)$ (see Eq.\eqref{boundary}).

Interestingly, for a particular form of the arguments, we can obtain the usual $f(R)$ gravity:
\begin{equation}\label{fR}
f(\Tw,\Bw)=f(-\Tw+\Bw)=f(\Rbol).
\end{equation}
It is precisely the boundary term $\Bw$ that can be used to link both $f(R)$ and $f(T)$ theories to the larger class of modified gravity theories $f(T,B)$ gravity, where $f(R)$ and $f(T)$ are now limiting cases.  A side result which emerges from studying $f(T,B)$ gravity models is that there is no direct link between $f(R)$ gravity and $f(T)$ gravity.

The case \eqref{fR} is particularly interesting in light of the discussion of the previous sections where we have addressed the issue of determining the spin connection corresponding to the tetrad and the pure tetrad formulation of teleparallel theories. From  the equivalence with $f(R)$ gravity, it is obvious that we do not face these problems as the only variable in $f(R)$ gravity is the metric tensor. Nevertheless, it is illustrative to understand this fact within the teleparallel framework.  As it turns out, the spin connection entering the boundary term in \eqref{fR} exactly cancels out the boundary term contribution of the spin connection to the $\Tw$-term from the relation \eqref{rel}.  This provides us with a $f(-\Tw+\Bw)$ term entirely independent of the spin connection and hence leads to  symmetric field equations. Therefore, the case \eqref{fR} is a teleparallel theory that is locally Lorentz invariant even in the pure tetrad formulation discussed in section~\ref{secPT}.

Recently, a number of other teleparallel theories with higher order field equations were proposed. These include, for instance,  the so-called teleparallel Gauss-Bonnet gravity \cite{Kofinas:2014owa,Bahamonde:2016kba} inspired by analogous work in the curvature approach \cite{Nojiri:2005jg}, where the following Lagrangian was considered
\begin{equation}
\Lw_{\rm fTT_G} =
\frac{h}{2\kappa} f(\Tw,\Tw_G).
\label{actionfTTG}
\end{equation}
Here $\Tw_G$ is the so-called teleparallel Gauss-Bonnet term related to the usual curvature Gauss-Bonnet by a boundary term in a similar fashion to the relationship of curvature and torsion scalars in Eq.\eqref{boundary}. Considering   a $f$-function of such a teleparallel Gauss-Bonnet term in \eqref{actionfTTG} leads  then to a distinctive gravity model to curvature Gauss-Bonnet gravity.
It is also possible to introduce derivatives of torsion in other ways (see~\cite{Otalora:2016dxe}).

\section{Final Remarks}

Teleparallel gravity and its generalizations can be formulated as fully invariant (both coordinate and local Lorentz) theories of gravity. Nevertheless, it is often suggested in the literature that torsion is not a tensor in teleparallel gravity or likewise that the local Lorentz symmetry is violated and teleparallel gravity theories are frame dependent.  These notions originated from the fact that the teleparallel spin connection  is of a pure-gauge form and hence it is always possible to choose a special gauge in which it vanishes. This is similar to choosing a specific gauge in gauge theories. This non-covariant approach where one restricts the analysis to a vanishing connection is what has been coined \emph{pure tetrad teleparallel gravity}. One of the primary goals of this paper is to distinguish between what happens in the pure tetrad and the invariant formulations of  teleparallel gravity and their generalizations, clearly illuminating the properties of the Lorentz connections and their pivotal role in understanding and determining the equations in Lorentz invariant gravitational theories.

In the invariant framework described here for teleparallel theories of gravity, the torsion tensor is a covariant object under both diffeomorphisms and local Lorentz transformations. However, unlike the familiar situation in general relativity where the curvature tensor depends only on the metric tensor, the torsion tensor of teleparallel gravity is a function of both the tetrad and the spin connection. The teleparallel spin connection is independent of the metric tensor and represents only the inertial effects associated with the choice of the frame. This is the crucial difference when comparing teleparallel theories with general relativity and other curvature-based theories of gravity, which introduces the very pressing practical problem of how to determine  both the tetrad and the teleparallel spin connection.

In the teleparallel equivalent of general relativity the teleparallel spin connection does not enter the field equations for the tetrad, and the field equations  for the spin connection turn out to be identically satisfied. Both of these properties can be easily understood as a consequence of the spin connection contributing to the teleparallel action through the surface term only. Therefore, as far as the solutions of the field equations are concerned, the spin connection can be chosen arbitrarily in order to solve the field equations. In particular, this allows us to set the spin connection to zero and effectively obtain the purely tetrad formulation of teleparallel gravity. 

However, in revealing  the underlying problem of the pure tetrad formulation, the crucial point  is that the spin connection can be chosen arbitrarily only when we are interested in solutions of the field equations. The spin connection still plays an important role as it contributes to the action through the surface term  which  manifests itself in many situations where the total value of the action is of interest; e.g., calculations of the energy-momentum and black hole thermodynamics. As was shown in~\cite{Krssak_Pereira2015}, in order to determine the spin connection corresponding to the tetrad, we can use the fact that the spin connection regularizes the action and hence we can define it by the requirement of the finiteness of the action. From a physical perspective, this amounts to the removal of spurious inertial contributions  causing divergences of the action and obtaining a purely gravitational action.  In practice, this can be achieved by introducing a reference tetrad which represents the same inertial effects as the full tetrad. This leads to the procedure described in Section~\ref{seccon}, which was  demonstrated explicitly for the spherically symmetric solution in Section~\ref{secExample}. It should be mentioned that this is the simplest, but not the only method to determine the spin connection. For example, see the recently developed  method  of determining the spin connection using the spacetime symmetries~\cite{Hohmann:2019nat}.

Note that the problem of how to determine the spin connection corresponding to the tetrad arises only if our starting point is the Lagrangian \eqref{TeleLagra} depending on both the tetrad and the spin connection as \textit{a priori} independent variables. The procedure discussed above can then be viewed as a method of determining their mutual relation. However, if we follow the gauge construction reviewed in Section~\ref{sec:GaugeStruct},  we naturally avoid this problem. We start with an inertial frame together with a gauge translational potential and a gauge covariant derivative which can be naturally introduced. The spin connection appears when we pass to the general frame by performing a local Lorentz transformation, in a similar fashion  to how it appears in special relativity. We can then see that the general tetrad \eqref{TeleTetrada} is given by a combination of the spin connection and the translational gauge potential and that the translational field strength \eqref{tfs} coincides with the torsion tensor \eqref{tfs2} for the general tetrad. In this construction, it is obvious that the tetrad is not an independent variable from the spin connection and that, in fact, to each tetrad corresponds some teleparallel spin connection. 

It is useful to remember that the field equations of teleparallel gravity are non-linear coupled PDEs that can only be solved analytically in certain highly symmetric situations. Therefore, as in the case in general relativity, a starting point of many explicit calculations is a certain ansatz which respects the assumed symmetry. From this ansatz metric  we then  choose  an ansatz tetrad and solve the field equations. This is the reason why in all practical calculations we start with the tetrad and the spin connection as \textit{a priori} independent variables instead of constructing the tetrad from the translational gauge potential and the spin connection. This approach of solving the field equations is   very much within the spirit of general relativity and it remains an open question as to whether one could  fully follow the  gauge construction  in practice and use translational gauge potentials instead of the tetrad (and  whether  there would be any advantage to such an approach). 

The situation is radically different in the case of modified teleparallel theories of gravity where the teleparallel spin connection contributes to the action in a more intricate way. The variation with respect to the tetrad and the spin connection results in a system of coupled field equations that depend on both variables in a non-trivial way. However, it turns out that  the resulting  field equations for the spin connection are equivalent to the antisymmetric part of the field equations for the tetrad. This means that, unlike in the case of the ordinary teleparallel gravity, there is no freedom to choose the spin connection when solving the field equations; instead the spin connection is determined by the field equations. Therefore, the solution to the problem in modified teleparallel theories is always the  pair $h^a{}_{\mu}$ and $\omw^a{}_{b\mu}$. It is interesting to note that in many highly symmetric situations, such as spherically symmetric spacetimes or isotropic cosmologies, the antisymmetric and symmetric parts of the field equations for the tetrad do decouple from each other~\cite{Krssak_Saridakis2015,HJKP}. As a result, it is often possible to solve the antisymmetric part of the field equations, and hence determine the spin connection, independently from obtaining the solution for the symmetric part of the field equations that determines the full tetrad and the metric tensor. 

We can now clearly understand the problem of the pure tetrad formulation in the modified case and why these theories can easily be misunderstood regarding their local Lorentz invariance. Since the solution of the field equations is always the pair $h^a_{\ \mu}$ and $\omw^a{}_{b\mu}$, the field equations in the pure tetrad formulation are non-trivially satisfied only in the case when the tetrad corresponds to a vanishing spin connection. These tetrads were originally nicknamed \textit{good tetrads} in the case of $f(T)$ gravity, since they lead to non-trivial solutions of the field equations; this is in contrast with the so-called \textit{bad tetrads}, in which case $f(T)$ gravity reduces trivially to ordinary teleparallel gravity~\cite{Tamanini_Boehmer2012}. We  now  see that this concept of good and bad tetrads is just the result of neglecting the role of the teleparallel spin connection and one cannot draw any conclusions about the preferred frames in teleparallel theories. Nevertheless, we should mention that despite these conceptual and fundamental flaws, the pure tetrad formulation -- if one properly uses good tetrads only -- can be utilized to successfully solve the field equations. Therefore, most of the results found in the literature -- obtained using the pure tetrad formulation -- are correct.

In Section~\ref{sec:Other-Modified}, we reviewed the covariant formulation of other modified teleparallel theories and classified various models based on their essential features. With the exception of higher derivative theories, equivalent to some curvature based models as discussed in Section~\ref{secModHD}, much of the previous discussion about the tetrad and the spin connection from $f(T)$ gravity generally applies to all modified teleparallel models~\cite{Hohmann:2017duq}.

It is worth noting that alternatively one could set up an action for teleparallel theories and through the use of Lagrange multipliers ensure that the curvature of the spin connection is zero via a metric affine gauge approach~\cite{Obukhov_Pereira2003,Hehl_McCrea_Mielke_Neeman1995}. In this case the action would be a functional of both the frame field and the spin connection. The result of varying with respect to the frame field will yield a set of equivalent field equations to the covariant version presented here and in~\cite{Krssak_Saridakis2015}. The result of varying with respect to the spin connection could, of course, result in a difference in the matter sector of the theory unless additional assumptions are placed on the nature of the matter Lagrangian (such as, for example, independence from the spin connection). Assuming that there is no hyper-momentum or spin current matter source~\cite{Hehl_McCrea_Mielke_Neeman1995}, the zero curvature constraint results in a Lorentz connection as in the covariant representation of the theory.

An interesting and very fundamental open problem is the question of the propagating degrees of freedom in modified teleparallel theories of gravity. In the case of $f(T)$ gravity it was shown that there are no propagating extra degrees of freedom at the linear level~\cite{Chen:2010va,Izumi:2012qj,Bamba:2013ooa}. However, the Hamiltonian analysis revealed that there are five propagating degrees of freedom~\cite{Li:2011rn}. It was then argued that the presence of these extra non-perturbative degrees of freedom poses a serious problem for the causality of $f(T)$ gravity~\cite{Ong:2013qja}, and these claims were then further discussed in~\cite{Izumi:2013dca, Ferraro_Fiorini2014,Chen:2014qtl}. Recently, these results were questioned and it was argued that $f(T)$ gravity has only one extra propagating degree of freedom~\cite{Ferraro:2018tpu,Ferraro:2018axk}, the same number as $f(R)$ gravity. Most of these results were obtained using the non-covariant pure tetrad teleparallel gravity, and therefore their applicability within the invariant framework presented here is not yet clear, see also the interesting discussion in~\cite{Bejarano:2019fii}. However, the recent analysis of the covariant formulation of new  general relativity, discussed in Section~\ref{sec:NGR}, implies that the presence  of the spin connection does not influence the total number of degrees of freedom~\cite{Hohmann:2018jso}. Similarly, a general analysis of teleparallel theories using the method of Lagrange multipliers led to the same conclusion~\cite{Nester:2017wau}. Nevertheless, this is clearly an important open question for future consideration.

We can mention here also a distinct class of theories introduced by Nester and Yo~\cite{Nester:1998mp}, where gravity is attributed to  the non-metricity of spacetime. Since the teleparallel condition of zero curvature is satisfied and the connection is symmetric due to vanishing torsion, this approach was named \textit{symmetric teleparallel gravity}. Within this framework it is possible to formulate the symmetric teleparallel equivalent of general relativity~\cite{Nester:1998mp,Adak:2005cd,Adak:2006rx,BeltranJimenez:2017tkd,BeltranJimenez:2018vdo}, which can be understood also as a gauge theory for translations~\cite{Koivisto:2018aip,Adak:2018vzk}, and construct new modified gravity models~\cite{Conroy:2017yln,Jarv:2018bgs,Runkla:2018xrv,Hohmann:2018wxu,Iosifidis:2018zwo}.

Let us conclude with the statement that teleparallel theories of gravity, which have experienced a renaissance recently, are an intriguing approach to understand gravity. In the case of ordinary teleparallel gravity, we are able to obtain a number of fundamental insights into the nature of gravity, which are not readily available (or are, at least, more hidden) in standard general relativity. Among those which we have reviewed here include the manifest gauge nature of gravity, and new approaches to understand the problems of the definition of the gravitational energy-momentum and regularization of the gravitational action. There are a number of other areas where teleparallel gravity can improve our understanding of gravity. For instance, it has been argued that coarse graining of the gravitational field equations of general relativity might be more naturally achieved within a teleparallel frame formulation of gravity~\cite{Brannlund:2010rs,VanDenHoogen:2017nyy}. The ability to parallelly transport in a path independent manner facilitates an integration procedure useful for the development an averaged theory of gravity. 

We have also discussed a number of modified teleparallel gravity models within the covariant formulation. Since these theories are distinct new models of gravity, it is ultimately up to observations to discriminate between them. Nevertheless, a number of theoretical challenges arise. In particular, we have focused on the question of local Lorentz invariance, and we have clearly demonstrated that the question is resolved due to the existence of the \emph{inertial} Lorentz connection. 

\subsection*{Acknowledgements}
The authors are grateful to James N. Nester for valuable feedback. The work of MK was funded by the European Regional Development Fund through the Centre of Excellence \textit{TK 133 The Dark Side of the Universe}. RvdH is supported by the St.~Francis Xavier University Council on Research.  JGP is partially supported by CNPq, Brazil. AC is supported by NSERC of Canada.

 This article is partly based upon work from COST Action CA15117 (Cosmology and Astrophysics Network for Theoretical Advances and Training Actions), supported by COST (European Cooperation in Science and Technology).


\bibliographystyle{Style}
\bibliography{Tele-Parallel-Reference-file}

\end{document}